\documentclass[aps,twocolumn,showpacs,superscriptaddress]{revtex4}
\usepackage{graphicx}
\usepackage{amsmath}
\usepackage{amssymb}
\usepackage{color}
\usepackage{bbm}
\usepackage{bm}

\definecolor{gold}{rgb}{1,0.75,0}

\begin{document}
\title{Study of pair production in inhomogeneous two-color electric fields using the computational quantum field theory}
\author{Z. L. Li \footnote{Corresponding author: zlli@cumtb.edu.cn}}
\affiliation{School of Science, China University of Mining and Technology, Beijing 100083, China}
\author{C. Gong}
\affiliation{School of Science, China University of Mining and Technology, Beijing 100083, China}
\affiliation{State Key Laboratory for GeoMechanics and Deep Underground Engineering, China University of Mining and Technology, Beijing 100083, China}
\author{Y. J. Li \footnote{Corresponding author: lyj@aphy.iphy.ac.cn}}
\affiliation{School of Science, China University of Mining and Technology, Beijing 100083, China}
\affiliation{State Key Laboratory for GeoMechanics and Deep Underground Engineering, China University of Mining and Technology, Beijing 100083, China}
\date{\today}

\begin{abstract}
We first demonstrate theoretically that the computational quantum field theory is equivalent to the quantum kinetic theory for pair creation in a spatially homogeneous and time-dependent electric field, then verify numerically their equivalence for pair creation in one-dimensional time-dependent electric fields, and finally investigate detailedly the effects of the field frequency, spatial width, pulse duration, and relative phase on dynamically assisted Schwinger pair production in an inhomogeneous two-color electric field. It is found that the enhancement effect of pair creation is very sensitive to the field frequency and generally very obvious for a shorter field width, a longer pulse duration, and a relative phase of maximizing the field strength. These results can provide a significant reference for the optimal control theory of pair creation which aims to maximize the created pair yield within a given scope of field parameters.
\end{abstract}


\maketitle

\section{Introduction}\label{sec:sec1}

Vacuum pair production is a phenomenon that vacuum in the presence of a strong electric field is unstable and decays into electron-positron pairs \cite{Piazza2012,Xie2017}. It is one of the most fascinating predictions of quantum electrodynamics (QED), and has been studied with different methods since Dirac put forward the relativistic wave equation and predicted the existence of positrons \cite{Dirac}. Sauter \cite{Sauter} calculated the transmission coefficient through the energy gap of vacuum by exactly solving the Dirac equation in a static electric field. Heisenberg and Euler  \cite{Heisenberg} deduced the leading pair production rate in a weak electric field from the imaginary part of one-loop effective Lagrangian for spinor QED, and defined the critical field strength $E_{\mathrm{cr}}=m^2c^3/|q|$, where $m$ is the particle mass, $c$ is the speed of light, and $q$ is the particle charge ($\hbar=1$ is used throughout the paper). Schwinger \cite{Schwinger} formalized pair creation process in the language of QED and recovered the pair production rate in a constant field with the proper-time method, so sometimes the tunneling pair creation is also called Schwinger pair production or Schwinger effect.

In addition to above-mentioned methods, the current widely used ones include the semiclassical scattering method related to the Wentzel-Kramers-Brillouin approach \cite{Brezin1970,Popov1972,Kim2007,Dumlu1011,Strobel2015,Oertel2019}, the worldline instanton technique \cite{Dunne2005,Dunne2006,Dumlu2011,Ilderton2015,Schneider2018}, the quantum Vlasov equation (QVE) \cite{Kluger1998,Schmidt1998,Bloch1999,Kohlfurst2014,Gong2020}, the Wigner function \cite{Bialynicki1991,Hebenstreit2010,Hebenstreit2011,ZLLi1517,Kohlfurst1819,Zlli2019}, the computational quantum field theory (CQFT) \cite{Krekora2004,Krekora2006,QSu2012,Gong2018,QSu2019} and so on. Note that the QVE and the Wigner function can be collectively called quantum kinetic theory (QKT), and the Dirac-Heisenberg-Wigner (DHW) formalism is the Wigner function in spinor QED. In recent years, the relation between these methods has been explored, because it can not only be used to mutually authenticate each other's results, but also help to understand their results from different perspectives. For instance, Dumlu \cite{Dumlu2009} proved the equivalence between the QVE and the semiclassical scattering method in both scalar QED and spinor QED. Hebenstreit \textit{et al}. \cite{Hebenstreit2010} found that the Wigner function in spinor QED could be reduced to the QVE for a spatially homogeneous and time-dependent electric field with one component. Li \textit{et al}. \cite{Zlli2019} generalized the above result to the scalar QED, and found that in this case the electric field could have three components. Blinne and Strobel \cite{Blinne2016} compared the semiclassical scattering method with the DHW formalism for rotating fields, and found that the numerical methods of these two approaches are complementary in terms of computation time as well as accuracy. Strobel and Xue \cite{Strobel2014} calculated pair creation rate for time-dependent electric fields with more than one component in scalar QED using the semiclassical scattering method and the worldline instanton technique, and found that the results obtained from these two methods agree with each other by expanding the momentum spectrum around zero point of the canonical momentum. A similar work was done in \cite{SPKim2019}. More recently, Unger \textit{et al}. \cite{Unger2019} explored the relation between the QVE and the CQFT, but found that they were not exactly equivalent. In this paper, We will focus on proving the equivalence between these two methods.

Based on the Schwinger formula, the pair creation rate $\sim\exp(-\pi E_\mathrm{cr}/E)$, to create observable electron-positron pairs the electric field strength $E$ should be comparable with the critical field strength $\sim10^{16}\mathrm{V/cm}$, corresponding to laser intensity $\sim10^{29}\mathrm{W/cm^2}$. However, the laser intensity in forthcoming laser facilities \cite{ELI,XCELS} is expected to reach $10^{24}-10^{26}\mathrm{W/cm^2}$, which is still much less than the critical one. So various methods of enhancing pair production are explored \cite{Schutzhold2008,Bell2008,Dunne2009,Piazza2009,Bulanov2010,Titov2012,ZLLi2014,Torgrimsson1619,Olugh2019}. One of the most effective mechanisms is the dynamically assisted Schwinger pair production (DASPP), which shows that the pair yield can be greatly enhanced by superposing a strong but slowly varying Sauter pulse with a weak but rapidly changing one. Orthaber \textit{et al}. \cite{Orthaber2011} further verified this mechanism employing the QVE instead of the world-line instanton technique and given the momentum spectra for combined electric fields. Nuriman \textit{et al}. \cite{Nuriman2012} studied the DASPP for different combinations of given electric fields and put forward the idea of optimizing pair production, which leads to the appearance of an optimal control theory for time-dependent electric fields \cite{Kohlfurst2013}. Based on the QVE, Otto \textit{et al}. \cite{Otto2015} provide a qualitative understanding of the DASPP in a bifrequent electric field with spatial homogeneity. Moreover, although the effect of spatially inhomogeneity on the DASPP has been more or less considered in \cite{Jiang2012,Schneider2016,Ababekri2019}, the fields are too simple to supply sufficient information to an optimal theory. In this paper, we will study in detail the effects of the field frequency, field width, pulse duration, and relative phase on the DASPP for a more complex but realistic field, i.e., an inhomogeneous two-color electric field, and the results may provide a great amount of firsthand information for perfecting a tentative optimal theory of pair creation in space-time dependent fields shown in \cite{Dong2020}.


This paper is structured as follows: In Sec. \ref{sec:sec2} the CQFT is introduced and compared with the QKT for pair creation in time-dependent electric fields with or without spatial inhomogeneity. The equivalence between these two methods is proved. In Sec. \ref{sec:sec3} the effects of the field frequency, field width, pulse duration, and relative phase on the DASPP for an inhomogeneous two-color electric field are investigated in detail. In Sec. \ref{sec:sec4} the conclusions and discussions are given. To clearly compare the CQFT with the QKT, the QVE and DHW formalism in QED$_{1+1}$ are shown in Appendixes \ref{app:Appa} and \ref{app:Appb}.

\section{Theoretical methods and their equivalence}\label{sec:sec2}

The starting point is the Dirac equation with $\hbar=1$:
\begin{equation}\label{eqn:DiracEquation1}
\{i\gamma^{\mu}[\partial_\mu+i q A_\mu(\mathbf{x},t)]-m c\}\Psi(\mathbf{x},t)=0,
\end{equation}
where $\partial_\mu=(\partial_t/c, \bm{\nabla})$, $q$ and $m$ are the particle charge and mass, respectively, $A_\mu(\mathbf{x},t)=(\varphi/c, -\mathbf{A}/c)$ is the four-potential of electromagnetic fields. The Dirac matrices $\gamma^{\mu}$ are represented in Dirac basis as
\begin{equation}
\gamma^0=\left(
           \begin{array}{cc}
            \mathbbm{1}_2 & 0 \\
             0 & -\mathbbm{1}_2 \\
           \end{array}
         \right), \quad
\bm{\gamma}=\left(
           \begin{array}{cc}
             0 & \bm{\sigma} \\
             -\bm{\sigma} & 0 \\
           \end{array}
         \right), \quad
\end{equation}
where $\mathbbm{1}_2$ is the $2\times2$ identify matrix and $\bm{\sigma}=(\sigma_x, \sigma_y, \sigma_z)$ are the Pauli matrices with
\begin{equation}
\sigma_x=\left(
           \begin{array}{cc}
             0 & 1 \\
             1 & 0 \\
           \end{array}
         \right), \quad
\sigma_y=\left(
           \begin{array}{cc}
             0 & -i \\
             i & 0 \\
           \end{array}
         \right), \quad
\sigma_z=\left(
           \begin{array}{cc}
             1 & 0 \\
             0 & -1 \\
           \end{array}
         \right).\nonumber
\end{equation}

\subsection{Equivalence between the CQFT and QKT for time-dependent electric fields}\label{sec:subseca}

In this subsection, we will prove in detail the equivalence between the CQFT and the QVE in spinor QED for an uniform and time-dependent electric field with one component, and show that the CQFT is equivalent to the QKT for arbitrary spatially homogeneous and time-dependent electric fields.

Equation (\ref{eqn:DiracEquation1}) can also be written as
\begin{equation}\label{eqn:DiracEquation2}
i\frac{\partial}{\partial t}\Psi(\mathbf{x},t)=\mathcal{H}(\mathbf{x},t)\Psi(\mathbf{x},t),
\end{equation}
where the time-dependent Hamiltonian
\begin{equation}
\mathcal{H}(\mathbf{x},t)=c \bm{\alpha}\cdot\big[\mathbf{P}-\frac{q}{c}\mathbf{A}(\mathbf{x},t)\big]+\beta mc^2+q\varphi(\mathbf{x},t),
\end{equation}\label{eqn:Hamiltonian}
$\mathbf{P}=-i\bm{\nabla}$ is the canonical momentum operator, $\bm{\alpha}=\gamma^0\bm{\gamma}=\left(
           \begin{array}{cc}
             0 & \bm{\sigma} \\
             \bm{\sigma} & 0 \\
           \end{array}
         \right)$ and $\beta=\gamma^0$.

For a spatially homogeneous and time-dependent electric field with temporal gauge $A_\mu(\mathbf{x},t)=(0,-\mathbf{A}(t)/c)=(0,0,0,-A_z(t)/c)$ and $E_z(t)=-dA_z(t)/dt$, the canonical momentum $\mathbf{k}$ is a good quantum number and the Dirac field can be decomposed as
\begin{equation}\label{eqn:FourierDeco}
\Psi(\mathbf{x},t)=\int\frac{d^3k}{(2\pi)^3}\Psi_\mathbf{k}(t)e^{i\mathbf{k}\cdot\mathbf{x}},
\end{equation}
where $\Psi_\mathbf{k}(t)$ are the Fourier modes.
Inserting the above equation into Eq. (\ref{eqn:DiracEquation2}), we have
\begin{equation}\label{eqn:DEinPS}
i\frac{\partial}{\partial t}\Psi_\mathbf{k}(t)=[c \bm{\alpha}\cdot\mathbf{p}(t)+\beta mc^2]\Psi_\mathbf{k}(t)=\mathcal{H}_\mathbf{p}(t)\Psi_\mathbf{k}(t),
\end{equation}
where $\mathbf{p}(t)=\mathbf{k}-q\mathbf{A}(t)/c$ is the kinetic momentum and $\mathcal{H}_\mathbf{p}(t)=$
\begin{equation}
\left(
                     \begin{array}{cccc}
                       m c^2 & 0 & c p_z & c (p_x-ip_y) \\
                       0 & m c^2 & c (p_x+ip_y) & -c p_z \\
                       c p_z & c (p_x-ip_y) & -m c^2 & 0 \\
                       c (p_x+ip_y) & -c p_z & 0 & -m c^2 \\
                     \end{array}
                   \right)\nonumber
\end{equation}
is the matrix form of the time-dependent Hamiltonian in momentum space whose eigenvalues and eigenvectors are $\pm \omega_\mathbf{p}=\pm [c^2\mathbf{p}^2+m^2c^4]^{1/2}$ and
\begin{equation}\label{eqn:eigenvectors}
\begin{split}
\mathcal{U}_{\mathbf{p},1}&=\frac{1}{\sqrt{4\omega_\mathbf{p}(\omega_\mathbf{p}-c p_z)}}\left(
     \begin{array}{c}
       \omega_\mathbf{p}+mc^2-c p_z \\
       -c(p_x+ip_y) \\
       -\omega_\mathbf{p}+mc^2+c p_z \\
       c(p_x+ip_y) \\
     \end{array}
   \right)\!,\;\;\\
\mathcal{U}_{\mathbf{p},2}&=\!\frac{1}{\sqrt{4\omega_\mathbf{p}(\omega_\mathbf{p}-c p_z)}}\left(
     \begin{array}{c}
       c(p_x-ip_y) \\
       \omega_\mathbf{p}+mc^2-c p_z \\
       c(p_x-ip_y) \\
       \omega_\mathbf{p}-mc^2-c p_z \\
     \end{array}
   \right)\!,\, \\
\mathcal{V}_{\mathbf{p},1}&=\frac{1}{\sqrt{4\omega_\mathbf{p}(\omega_\mathbf{p}+c p_z)}}\left(
     \begin{array}{c}
       -\omega_\mathbf{p}+mc^2-c p_z \\
       -c(p_x+ip_y) \\
       \omega_\mathbf{p}+mc^2+c p_z \\
       c(p_x+ip_y) \\
     \end{array}
   \right)\!,\;\;\\
\mathcal{V}_{\mathbf{p},2}&=\!\frac{1}{\sqrt{4\omega_\mathbf{p}(\omega_\mathbf{p}+c p_z)}}\left(
     \begin{array}{c}
       c(p_x-ip_y) \\
       -\omega_\mathbf{p}+mc^2-c p_z \\
       c(p_x-ip_y) \\
       -\omega_\mathbf{p}-mc^2-c p_z \\
     \end{array}
   \right),
\end{split}
\end{equation}
where $\mathcal{U}_{\mathbf{p},r}^\dagger \mathcal{U}_{\mathbf{p},s}=\mathcal{V}_{\mathbf{p},r}^\dagger \mathcal{V}_{\mathbf{p},s}=\delta_{rs}$, $\mathcal{U}_{\mathbf{p},r}^\dagger \mathcal{V}_{\mathbf{p},s}=\mathcal{V}_{\mathbf{p},r}^\dagger \mathcal{U}_{\mathbf{p},s}=0$, and $r,s=\{1,2\}$. Note that the instantaneous eigenvectors of the time-dependent Hamiltonian are calculated by a numerical method in the CQFT \cite{Krekora2006}.


The Dirac field operator at time $t$ in momentum space can be obtained by evolving the field operator of field-free Dirac Hamiltonian $\Psi_\mathbf{k}(t_0)$ as
\begin{equation}\label{eqn:Evolution}
\Psi_\mathbf{k}(t)=U(t,t_0)\Psi_\mathbf{k}(t_0),
\end{equation}
where $U(t,t_0)=\mathcal{T}\exp(-i{\int^t_{t_0}\mathcal{H}_\mathbf{p}(t')dt'})$ is the time-evolution operator and $\mathcal{T}$ is the time-ordering operator. Accordingly, the positive and negative energy states $u_{\mathbf{k},s}(t)e^{i\mathbf{k}\cdot\mathbf{x}}$ and $v_{-\mathbf{k},s}(t)e^{i\mathbf{k}\cdot\mathbf{x}}$, which are the solutions of Eq. (\ref{eqn:DEinPS}), can be calculated by the time evolution of the force-free positive and negative energy eigenstates $u_{\mathbf{k},s}(t_0)e^{i\mathbf{k}\cdot\mathbf{x}}=\mathcal{U}_{\mathbf{p},s}(t_0)e^{i(\mathbf{k}\cdot\mathbf{x}-\omega_\mathbf{k}t_0)}$ and $v_{-\mathbf{k},s}(t_0)e^{i\mathbf{k}\cdot\mathbf{x}}=\mathcal{V}_{\mathbf{p},s}(t_0)
e^{i(\mathbf{k}\cdot\mathbf{x}+\omega_\mathbf{k}t_0)}$ using the numerical split-operator technique \cite{Braun1999,Mocken2008}, i.e., $u_{\mathbf{k},s}(t)=U(t,t_0)u_{\mathbf{k},s}(t_0)$ and $v_{-\mathbf{k},s}(t)=U(t,t_0)v_{-\mathbf{k},s}(t_0)$.

Therefore, the field operator can be expanded in terms of  the positive and negative energy states and the time-independent particle annihilation operator $b_{\mathbf{k},s}$ and antiparticle creation operator $d_{-\mathbf{k},s}^{\dagger}$:
\begin{equation}\label{eqn:FO1}
\Psi(\mathbf{x},t)=\!\int\!\frac{d^3k}{(2\pi)^3}\sum_{s=1}^2\big[b_{\mathbf{k},s}u_{\mathbf{k},s}(t)
+d_{-\mathbf{k},s}^{\dagger}v_{-\mathbf{k},s}(t)\big]e^{i\mathbf{k}\cdot\mathbf{x}},
\end{equation}
where $b_{\mathbf{k},s}|\mathrm{vac}\rangle=\langle\mathrm{vac}|d_{-\mathbf{k},s}^{\dagger}=0$, and the creation and annihilation operators fulfil the standard fermionic anticommutation relations.
Furthermore, the field operator can also be expressed by the instantaneous eigenvectors of the Hamiltonian as
\begin{equation}\label{eqn:FO2}
\begin{split}
\Psi(\mathbf{x},t)=\int\frac{d^3k}{(2\pi)^3}\sum_{s=1}^2\big[&\mathcal{B}_{\mathbf{k},s}(t)
\,\mathcal{U}_{\mathbf{k},s}(t)\\
&+\mathcal{D}_{-\mathbf{k},s}^{\dagger}(t)\mathcal{V}_{-\mathbf{k},s}(t)\big]
e^{i\mathbf{k}\cdot\mathbf{x}},
\end{split}
\end{equation}
where $\mathcal{B}_{\mathbf{k},s}(t)$ and $\mathcal{D}_{-\mathbf{k},s}^{\dagger}(t)$ are the time-dependent particle annihilation operator and antiparticle creation operator, respectively, which satisfy the standard fermionic anticommutation relations as well. The expression Eq. (\ref{eqn:FO2}) can clearly distinguish the positive and negative energy states in the presence of external fields and give a meaningful interpretation of particles and antiparticles, because the Hamiltonian is diagonal. However, since the external fields are nonvanishing, the created particles should be understood as quasi-particles. The total number density of created electrons is defined as
\begin{equation}\label{eqn:Defination}
\mathcal{N}(\mathbf{x},t)=\sum_{s=1}^2\langle\mathrm{vac}|\Psi_{+,s}^\dagger(\mathbf{x},t)
\Psi_{+,s}(\mathbf{x},t)|\mathrm{vac}\rangle.
\end{equation}
where $\Psi_{+,s}(\mathbf{x},t)=\int\frac{d^3k}{(2\pi)^3}\mathcal{B}_{\mathbf{k},s}(t)\,\mathcal{U}_{\mathbf{k},s}(t)
e^{i\mathbf{k}\cdot\mathbf{x}}$ is the positive energy portion of the field operator.
From Eqs. (\ref{eqn:FO1}) and (\ref{eqn:FO2}), we find that
\begin{equation}\label{eqn:OperatorRelation}
\begin{split}
\mathcal{B}_{\mathbf{k},s}(t)&=b_{\mathbf{k},s}\mathcal{U}^\dagger_{\mathbf{k},s}(t)u_{\mathbf{k},s}(t)
+d_{-\mathbf{k},s}^{\dagger}\mathcal{U}^\dagger_{\mathbf{k},s}(t) v_{-\mathbf{k},s}(t),\\
\mathcal{D}^\dagger_{-\mathbf{k},s}(t)&=b_{\mathbf{k},s}\mathcal{V}^\dagger_{-\mathbf{k},s}(t)u_{\mathbf{k},s}(t)
+d_{-\mathbf{k},s}^{\dagger}\mathcal{V}^\dagger_{-\mathbf{k},s}(t) v_{-\mathbf{k},s}(t).
\end{split}
\end{equation}
Then according to Eqs. (\ref{eqn:Defination}) and (\ref{eqn:OperatorRelation}), we obtain
\begin{equation}\label{eqn:NumberDensity0}
\mathcal{N}(\mathbf{x},t)=\mathcal{N}(t)=\int\frac{d^3k}{(2\pi)^3}\mathcal{F}_\mathbf{k}(t),
\end{equation}
where
\begin{equation}\label{eqn:MomentumDistribution}
\mathcal{F}_\mathbf{k}(t)=2\big|\mathcal{U}^\dagger_{\mathbf{k},s}(t) v_{-\mathbf{k},s}(t)\big|^2
\end{equation}
is the momentum distribution function. Note that here the spin index $s$ denotes a specific spin. Comparing with Eqs. (\ref{eqn:NumberDensity}) and (\ref{eqn:BogoliubovCoefficient}) in Appendix \ref{app:Appa}, one can clearly see that although there is a phase difference $\exp(-i\int^t_{t_0}\omega_\mathbf{k}(\tau)d\tau)$ between $\mathcal{U}^\dagger_{\mathbf{k},s}(t)$ and $\widetilde{u}^{\,\dagger}_{\mathbf{k},s}(t)$ [cf. Eqs. (\ref{eqn:eigenvectors}) and (\ref{eqn:FieldOperator40}) in Appendix \ref{app:Appa}], $\mathcal{F}_\mathbf{k}(t)$ is nothing but the one-particle momentum distribution function $F_\mathbf{k}(t)$ defined in the QVE, because the dynamical phase has no effect on the final result. Therefore, we finally confirm that the CQFT for a spatially homogeneous and time-dependent electric field with one component is completely equivalent to the QVE in spinor QED.

Furthermore, due to the equivalence between the QVE in spinor QED and the DHW formalism for a time-dependent electric field with one component \cite{Hebenstreit2010}, the CQFT in spinor QED is also equivalent to the DHW formalism for this field, which indicates that the definition of quasi-particles in the DHW formalism is based on instantaneous energy eigenstates as well. Similarly, one can verify the equivalence between the CQFT and the Wigner function in scalar QED, and generalize it to the time-dependent electric field with three components because for this field the QVE and the Wigner function in scalar QED are equivalent \cite{Zlli2019}. Based on these results and given that both the CQFT presented here and the DHW formalism shown in \cite{Blinne2014} can calculate the pair production in a time-dependent electric field with three components, the equivalence between the CQFT in spinor QED and the DHW formalism can be further generalized to time-dependent electric fields with three components. Though a direct theoretical proof of this generalization is difficult, a numerical verification is quite easy to achieve and we no longer show it here.

\subsection{Equivalence between the CQFT and QKT for one-dimensional time-dependent electric fields}\label{sec:subsecb}

In this subsection, we will show numerically that the CQFT is equivalent to the DHW formalism for a $1+1$ dimensional (one-dimensional space plus time) electric field.


Since the CQFT in the scalar potential gauge is equivalent to that in the vector potential gauge and the former case can save more computing time than the latter one \cite{Su2012}, we only show the CQFT in the scalar potential gauge here.

For a $1+1$ dimensional electric field $A_\mu(\mathbf{x},t)=(\varphi(z,t),\mathbf{0})$ where $E_z(z,t)=-\partial \varphi(z,t)/\partial z$, the Dirac equation (\ref{eqn:DiracEquation2}) becomes
\begin{equation}\label{eqn:DiracEquation3}
i\frac{\partial}{\partial t}\Psi(z,t)=\mathcal{H}(z,t)\Psi(z,t)
\end{equation}
with the one-dimensional time-dependent Hamiltonian $\mathcal{H}(z,t)=c \alpha_z P_z+\beta mc^2+q\varphi(z,t)$. Considering only a single spin direction, the four components of the quantum field operator can be reduced to only two, and the Hamiltonian is further reduced to
\begin{equation}\label{eqn:Hamiltonian1}
\mathcal{H}(z,t)=c \sigma_x P_z+\sigma_z mc^2+q\varphi(z,t).
\end{equation}

In the scalar potential gauge, the field operator is expanded in terms of the positive and negative energy states of the force-free Hamiltonian (In the following derivation, we omit the subscript $z$ and set $t_0=0$), $u_p(z)=\frac{e^{ipz}}{2\pi\sqrt{2\omega_p}}\left(\!
                                  \begin{array}{cc}
                                    \sqrt{\omega_p+mc^2} \;&
                                    \mathrm{sgn}(p)\sqrt{\omega_p-mc^2} \\
                                  \end{array}
                                \right)^\textsf{T}
$ and $v_n(z)=\frac{e^{inz}}{2\pi\sqrt{2\omega_n}}\left(
                                  \begin{array}{cc}
                                    -\mathrm{sgn}(n)\sqrt{\omega_n-mc^2} \;&
                                    \sqrt{\omega_n+mc^2}\\
                                  \end{array}
                                \right)^\textsf{T}$,
as well as the positive and negative energy states $u_p(z,t)$ and $v_n(z,t)$ which are the time evolution of $u_p(z)$ and $v_n(z)$ under the whole Hamiltonian (\ref{eqn:Hamiltonian1}):
\begin{equation}\label{eqn:Fourier}
\begin{split}
\Psi(z,t)&=\sum_p b_p(t)u_p(z)+\sum_n d_n^{\dagger}(t)v_n(z)\\
&=\sum_p b_pu_p(z,t)+\sum_n d_n^{\dagger}v_n(z,t),
\end{split}
\end{equation}
where the only nonvanishing anticommutators of the creation and annihilation operators are $\{b_p(t),b^{\,\dagger}_{p'}(t)\}=2\pi\delta(p-p')$, $\{d_n(t),d^{\,\dagger}_{n'}(t)\}=2\pi\delta(n-n')$, $\{b_p,b^{\,\dagger}_{p'}\}=2\pi\delta(p-p')$, $\{d_n,d^{\,\dagger}_{n'}\}=2\pi\delta(n-n')$, and $b_p|\mathrm{vac}\rangle=\langle\mathrm{vac}|d_n^{\dagger}=0$. Thus,
\begin{equation}\label{eqn:Relationship}
\begin{split}
b_p(t)&=\sum_{p'} b_{p'}U_{p,p'}(t)+\sum_n d_n^{\dagger}U_{p,n}(t)\\
d_n^{\dagger}(t)&=\sum_p b_pU_{n,p}(t)+\sum_n d_n^{\dagger}U_{n,n'}(t),
\end{split}
\end{equation}
where the parameters $U_{p,p'}(t)=\int dz\,u_p^\dagger(z)u_{p'}(z,t)$, $U_{p,n}(t)=\int dz\,u_p^\dagger(z)v_{n}(z,t)$, $U_{n,p}(t)=\int dz\,v_n^\dagger(z)u_{p}(z,t)$, and $U_{n,n'}(t)=\int dz\,v_n^\dagger(z)v_{n'}(z,t)$ are the matrix elements of the time evolution operator which can be computed by solving the single-particle Dirac equation with the numerical split-operator technique. The probability density of created electrons is defined as
\begin{equation}\label{eqn:ProDen}
\begin{split}
\rho(z,t)&=\langle\mathrm{vac}|\Psi_{e}^\dagger(z,t)\Psi_{e}(z,t)|\mathrm{vac}\rangle \\
&=\sum_n\Big|\sum_pU_{p,n}(t)u_p(z)\Big|^2,
\end{split}
\end{equation}
where $\Psi_{e}(z,t)=\sum_p b_p(t)u_p(z)$ is the electron portion of the field operator. Then the average number of created electrons is obtained by integrating the probability density over space:
\begin{equation}\label{eqn:NumDen}
\begin{split}
N(t)&=\!\!\int\!dz\,\rho(z,t)=\sum_p\Big(\sum_n| U_{p,n}(t)|^2\Big)\\
&=\sum_p \rho(p,t),
\end{split}
\end{equation}
where $\rho(p,t)=\sum_n| U_{p,n}(t)|^2$ is called the momentum distribution function in the CQFT, but it is different from the definition of that in the DHW formalism $f(p,t)$, see Appendix \ref{app:Appb}.  The relation between these two definitions is $\rho(p,t)=f(p,t)/L$ with $L$ denoting the length of the numerical box.

To better show the equivalence between the CQFT and the DHW formalism for $1+1$ dimensional electric fields numerically, we consider a more complex external field
\begin{equation}
\begin{split}
E(z,t)=&E_0g(z)h(t) \\
=&E_0\,\mathrm{sech}^2\Big(\frac{z}{\lambda_0}\Big)\exp\Big(\!\!-\!\frac{t^2}{2\tau_0^2}\Big)\sin(\omega_0 t\!),
\end{split}
\end{equation}
where $E_0$ is the field amplitude, $\lambda_0$ denotes the field width, $\tau_0$ is the pulse duration, and $\omega_0$ is the field frequency.

\begin{figure*}[!ht]
  \centering
  \includegraphics[width=15cm]{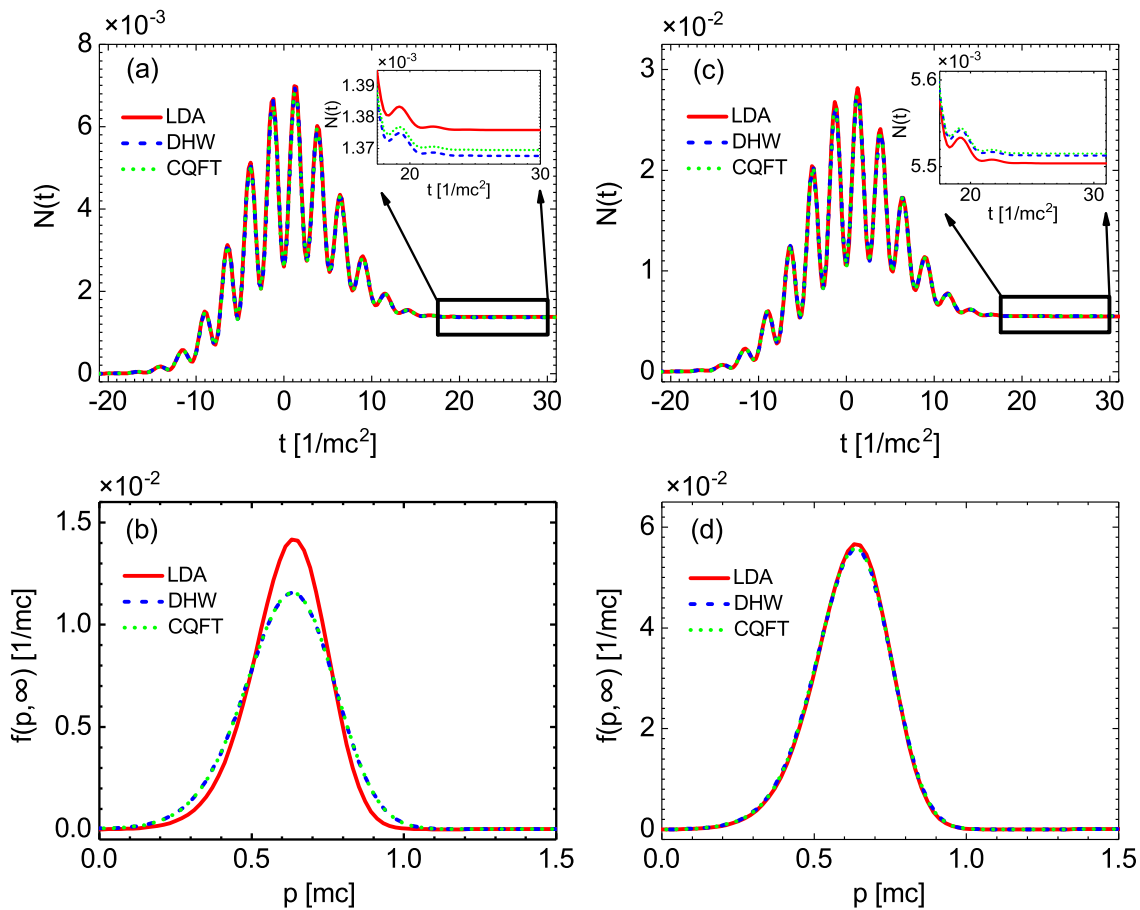}\\
  \caption{Pair yield evolving with time (upper row) and the momentum spectra of created particles (lower row) calculated by three different methods for $\lambda_0=8/mc$ (left column) and $\lambda_0=32/mc$ (right column). Other field parameters are $E_{0}=0.15625 E_\mathrm{cr}$, $\omega_0=1.2mc^2$, and $\tau_0=7.5/mc^2$.}\label{Fig1}
\end{figure*}

Figure \ref{Fig1} shows the comparisons of the time evolution of pair yield and the momentum spectra of created pairs calculated by three different approaches. The solid red line is the result of the local density approximation (LDA). The main idea of this approximation is that when the spatial variation scale of electric field is much larger than the Compton wavelength, the pair creation at any spatial point $z_i$ can be considered as occurring in a spatially homogeneous and time-dependent electric field $E(t)$ with field strength $E_0g(z_i)$. Therefore, we can calculate the momentum spectra and pair yield of created particles by solving the QVE repeatedly for the electric field $E(t)=E_0g(z_i)h(t)$ at any spatial point $z_i$ \cite{Hebenstreit2011}. However, it should be noticed that since the result calculated by the LDA is not the solution of Eqs. (\ref{eqn:DHW11})-(\ref{eqn:DHW12}), it can not give the correct particle number density in the coordinate space as well as the charge density. From the momentum spectra at $t\rightarrow\infty$, see Fig. 1(b) and (d), we find that the results of the CQFT and the DHW formalism coincide completely for both $\lambda_0=8/mc$ and $32/mc$. However, the result of the LDA is in good agreement with those of the other two methods for $\lambda_0=32/mc$, while is obviously different from them for $\lambda_0=8/mc$. Similar result can also be seen in Fig. \ref{Fig1}(a) and (c). Actually, the LDA is a very good approximation to study pair production in a quasi-homogeneous electric field whose spatial width divided by the speed of light is much greater than the pulse duration.

Although the momentum spectra at $t\rightarrow\infty$ given by the CQFT and the DHW formalism are the same, we are still not able to say that these two methods are completely equivalent, because the pair yield and the momentum spectra of created pairs obtained by expanding the Dirac field operator with different basis are the same when the external field is turned off but different when the field is still turned on \cite{Unger2019,Dabrowski2014}. Therefore, we further calculate the time evolution of pair yield with the CQFT and the DHW formalism, see Fig. \ref{Fig1}(a) and (c). One can see that the pair yield calculated by these two methods are the same at any time. This also holds true for momentum spectra. To differentiate the numerical results of the CQFT and the DHW formalism, we plot two insets in Fig. \ref{Fig1}(a) and (c), and find that the relative deviation between these two numerical results $|N_{\mathrm{CQFT}}-N_{\mathrm{DHW}}|/(N_{\mathrm{CQFT}}+N_{\mathrm{DHW}})$ is less than or equal to $0.1\%$. Of course this relative deviation can be further reduced by improving the numerical accuracy, because the difference of the numerical results between the CQFT and the DHW formalism is mainly caused by the accuracy of different numerical methods (the split-operator technique used in the CQFT and the spectral method used in the DHW formalism). So the CQFT is equivalent to the DHW formalism in QED$_{1+1}$ within an acceptable deviation. Furthermore, since in the CQFT with the vector potential gauge the Dirac field operator is expanded with instantaneous energy eigenstates, we can deduce that the definition of quasi-particles in the DHW formalism is also based on the instantaneous eigenstates.


\section{Numerical Results}\label{sec:sec3}
In this section, we mainly study the effect of field parameters on the dynamically assisted Schwinger pair production (DASPP). The external electric field we considered is a combination of a strong low-frequency field $E_\mathrm{s}(z,t)$ and a weak high-frequency field $E_\mathrm{w}(z,t)$, i.e.,
\begin{equation}
\begin{split}
E(z,t)=&E_\mathrm{s}(z,t)+E_\mathrm{w}(z,t)\\
=&E_\mathrm{s}\,\exp\Big(\!\!-\!\frac{z^2}{2\lambda_\mathrm{s}^2}\Big)
\exp\Big(\!\!-\!\frac{t^2}{2\tau^2}\Big)\cos(\omega_\mathrm{s} t)  \\
&+E_\mathrm{w}\,\exp\Big(\!\!-\!\frac{z^2}{2\lambda_\mathrm{w}^2}\Big)
\exp\Big(\!\!-\!\frac{t^2}{2\tau^2}\Big)\cos(\omega_\mathrm{w} t+\phi),
\end{split}
\end{equation}
where $E_{\mathrm{s,w}}$ are the field amplitudes, $\lambda_{\mathrm{s,w}}$ denote the field widths, $\tau$ is the pulse duration, $\omega_{\mathrm{s,w}}$ are the field frequencies, and $\phi$ is the relative phase of these two fields. The field amplitudes $E_\mathrm{s}=0.2E_\mathrm{cr}$ and $E_\mathrm{w}=0.05E_\mathrm{cr}$ are fixed. For the sake of analyzing the following results, the Keldysh adiabaticity parameter $\gamma=mc\omega/qE$ \cite{Keldysh1964} is introduced to distinguish between tunneling pair production ($\gamma\ll1$) and multiphoton pair creation ($\gamma\gg1$).

\subsection{Effect of the field frequency on DASPP}\label{sec:subsec3a}

\begin{figure}[!ht]
  \centering
  \includegraphics[width=8.5cm]{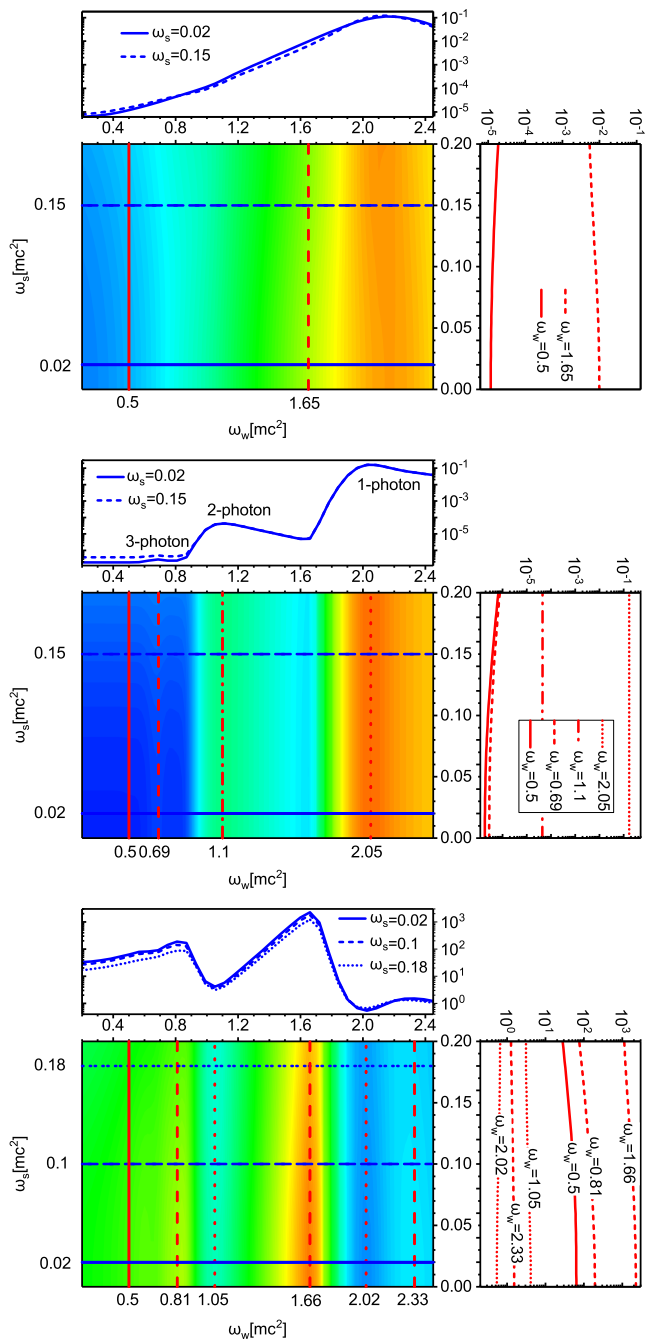}\\
  \caption{Contour profiles of the pair yield for combined electric fields $N_{\mathrm{s+w}}$ (top panel), the sum of the pair yield for strong fields and weak fields $N_\mathrm{s}$+$N_\mathrm{w}$ (middle panel), and the enhancement factor $N_{\mathrm{s+w}}/(N_\mathrm{s}+N_\mathrm{w})$ (bottom panel) varying with the field frequencies $\omega_{\mathrm{s,w}}$. Other field parameters are $\lambda_\mathrm{s}=\lambda_\mathrm{w}=10/mc$, $\tau=10/mc^2$, and $\phi=0$.}\label{Fig2}
\end{figure}

We depict the contour profiles of the pair yield for combined electric fields $N_{\mathrm{s+w}}$ (top panel), the sum of the pair yield for strong fields and weak fields $N_\mathrm{s}$+$N_\mathrm{w}$ (middle panel), and the enhancement factor $N_{\mathrm{s+w}}/(N_\mathrm{s}+N_\mathrm{w})$ (bottom panel) varying with the field frequencies $\omega_{\mathrm{s,w}}$ in Fig. \ref{Fig2}. The field parameters are chosen as $\lambda_\mathrm{s}=\lambda_\mathrm{w}=10/mc$, $\tau=10/mc^2$, and $\phi=0$. Note that we do not show the pair yield as a function of the field frequency for a single strong field and/or a single weak field, but these information can be obtained from the middle panel in Fig. \ref{Fig2}, because when one of the field frequencies $\omega_{\mathrm{s,w}}$ is fixed, the pair yield for the field corresponding to this frequency does not change with the other frequency, the trend of the relation between total pair yield and the other frequency is just that between pair yield and this frequency when only the corresponding field exists.

From the top panel in Fig. \ref{Fig2}, one can see that when the frequency of the strong field $\omega_\mathrm{s}$ is fixed and the frequency of the weak field $\omega_\mathrm{w}\leq2.15mc^2$, the pair yield for the combined electric field increases monotonically with the frequency of the weak field, which is different from the multiphoton peak structures on the relation curve between pair yield for only a single weak field and the field frequency, see the middle panel in Fig. \ref{Fig2}. This result can be understood by analyzing the combined Keldysh parameter $\gamma_\mathrm{c}=mc\omega_\mathrm{w}/qE_\mathrm{s}$. For a single weak field, it is found that the multiphoton peaks corresponding to $1$-, $2$-, $3$-photon pair production are present while the peaks corresponding to $\ell$-photon pair production with $\ell\geq4$ are absent, which indicates that only when the Keldysh parameter $\gamma_\mathrm{w}=mc\omega_\mathrm{w}/qE_\mathrm{w}\geq0.5/0.05=10\gg1$ ($0.5mc^2$ is a simple estimation of the field frequency corresponding to $4$-photon pair creation) does an obvious multiphoton peak appear. Therefore, there is no multiphoton peak except the one corresponding to $1$-photon pair creation at about $\omega_\mathrm{w}=2.15mc^2$ because the combined Keldysh parameter $\gamma_\mathrm{c}$ changes from $1$ to $12$. When $\omega_\mathrm{w}>2.15mc^2$, the pair yield decreases with increasing the field frequency of the weak high-frequency field, which is the same as the result obtained in \cite{Abdukerim2013}. In addition, the frequency of strong fields has little effect on the trend of the relation between pair yield and the frequency of weak fields.

As the frequency of weak fields $\omega_\mathrm{w}$ is fixed, the pair yield changes little with the frequency of strong fields $\omega_\mathrm{s}$ from $0$ to $0.2mc^2$, because the Keldysh parameter for strong fields $\gamma_\mathrm{s}=mc\omega_\mathrm{s}/qE_\mathrm{s}$ changes from $0$ to $1$, the contribution of strong fields to the pair creation is dominated by the field intensity rather than the frequency. Furthermore, the pair yield does not always increase with increasing the frequency of strong fields, it is also affected by the frequency of weak fields, and in some cases it decreases with the increase of the frequency of strong fields.

The bottom panel in Fig. \ref{Fig2} shows that the enhancement factor basically decreases with the frequency of strong fields, and oscillates with the frequency of weak fields. For the latter case, the maximum values of the enhancement factor appear at $\omega_\mathrm{w}=0.81,1.66,2.33mc^2$, particularly, the enhancement factor can reach about $2000$ at $1.66mc^2$, and can reach $200$ at $0.81mc^2$. The minimum values of the enhancement factor appear at $\omega_\mathrm{w}=1.05,2.02mc^2$, especially at $2.02mc^2$ the enhancement factor is $0.53<1$, which indicates that the increase of field frequency does not always enhance pair creation and sometimes reduces the pair yield. Furthermore, the points where the enhancement factor is minimal are exactly the frequencies corresponding to the multiphoton peaks on the relation curve between the pair yield and the frequency of weak fields. Intuitively, the reason is that the pair yield for combined fields increases monotonically with the small frequency of weak fields, the difference between the pair yield for the combined fields and the sum of pair yield for a single strong field and a single weak field at the points of multiphoton peaks is small.



\subsection{Effect of the field width on DASPP}\label{sec:subsec3b}

In this subsection, we mainly study the effect of the field width on DASPP in three different cases: $\lambda_\mathrm{s}=\lambda_\mathrm{w}$, $\lambda_\mathrm{s}>\lambda_\mathrm{w}$, and $\lambda_\mathrm{s}<\lambda_\mathrm{w}$.

\subsubsection{$\lambda_\mathrm{s}=\lambda_\mathrm{w}$}\label{sec:subsec3b1}

\begin{figure}[!ht]
  \centering
  \includegraphics[width=8.5cm]{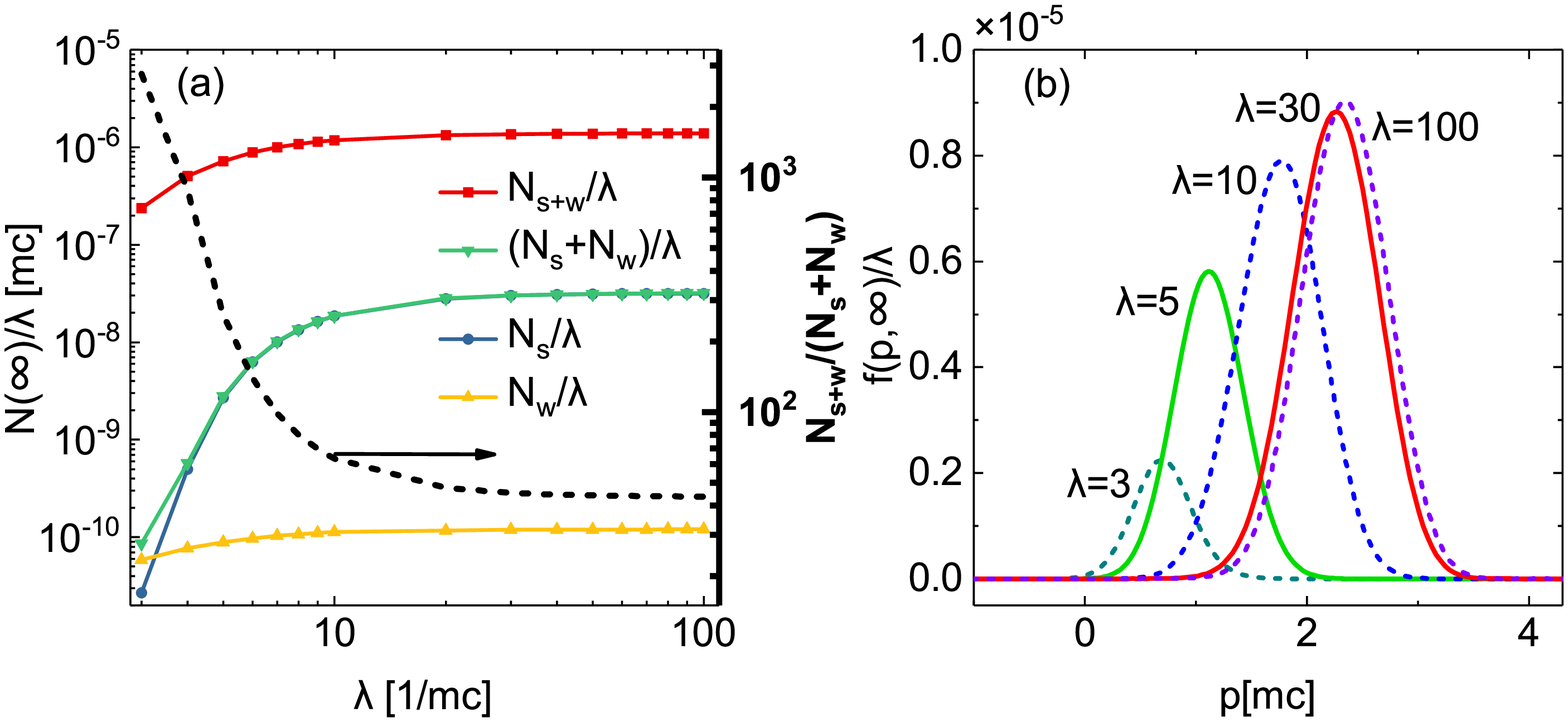}\\
  \caption{(a) Reduced pair yield for combined electric fields $N_\mathrm{s+w}/\lambda$ (red line and squares), strong fields $N_\mathrm{s}/\lambda$ (blue line and circles), weak fields $N_\mathrm{w}/\lambda$ (yellow line and triangles), the sum of the reduced pair yield for strong fields and weak fields $(N_\mathrm{s}+N_\mathrm{w})/\lambda$ (green line and inverted triangles), and the enhancement factor $N_{\mathrm{s+w}}/(N_\mathrm{s}+N_\mathrm{w})$ (black dashed line) as a function of the field width $\lambda$ (=$\lambda_\mathrm{s}$=$\lambda_\mathrm{w}$). The right arrow indicates the axis corresponding to the enhancement factor. (b) Reduced momentum spectra for the combined electric fields with $\lambda=3,5,10,30,100/mc$. Other field parameters are $\omega_\mathrm{s}=0.02mc^2$, $\omega_\mathrm{w}=0.5mc^2$,  $\tau=10/mc^2$, and $\phi=0$.}\label{Fig3}
\end{figure}

In this case, the reduced pair yield for combined electric fields $N_\mathrm{s+w}/\lambda$ (red line and squares), strong fields $N_\mathrm{s}/\lambda$ (blue line and circles), weak fields $N_\mathrm{w}/\lambda$ (yellow line and triangles), the sum of the reduced pair yield for strong fields and the weak fields $(N_\mathrm{s}+N_\mathrm{w})/\lambda$ (green line and inverted triangles), and the enhancement factor $N_{\mathrm{s+w}}/(N_\mathrm{s}+N_\mathrm{w})$ (black dashed line) as a function of the field width $\lambda$ (=$\lambda_\mathrm{s}$=$\lambda_\mathrm{w}$) are depicted in Fig. \ref{Fig3}. Note that since the total energy of electric fields, $\int\mathrm{d}z|E(z,t)|^2/2$, is proportional to the field width, the increase of the field width will increase the field energy and lead to the growth of pair yield. In order to eliminate this trivial effect, the reduced pair yield $N(\infty)/\lambda$ and momentum spectrum $f(p,\infty)/\lambda$ are used in the calculations.

It can be seen from Fig. \ref{Fig3}(a) that for combined fields the pair yield first increases rapidly and then linearly with the field width. The variation of pair yield with the field width for strong low-frequency fields is similar to that for the combined fields, but when the field width is small enough, the pair creation will decrease sharply with decreasing the field width and terminate. This is because for tunneling pair creation only the work done by the electric field force is greater than or equal to the vacuum energy gap $2mc^2$ that it can pull apart the virtual particle-antiparticle pair in vacuum to create a pair of real particles. For weak high-frequency fields, the pair yield changes gently with the increase of field width, and there is no the phenomenon that the pair yield drops dramatically with the field width and vanishes. The reason is that the pair production for a weak high-frequency field is dominated by multiphoton absorption process and can occur as long as the total energy of absorbed photons is greater than or equal to the energy gap $2mc^2$. Of course, this requires that the electric field can provide enough photons. Due to the fact that the number of photons decreases with decreasing the total energy of electric field, the multiphoton pair production process will also terminate once the electric field cannot provide enough photons.


The study of the enhancement factor finds that with the increase of field width the enhancement factor decreases from about $3000$ to about $40$, i.e., the enhancement effect of pair creation becomes weaker and weaker with the increase of field width. In fact, further study shows that when the field width is small enough, the enhancement factor will also decrease with decreasing the field width, and there will be one or two maximum values near the thresholds of field width corresponding to strong fields and weak fields. Notice here that the threshold of field width is the critical value of field width determining the initiation and termination of pair production. Since pair creation process near the threshold is very sensitive to the change of total energy of electric fields, and for a smaller field width the pair yield decreases sharply with the decrease of the field width and finally vanishes, the increase of electric field energy near the threshold of field width can restart the terminated pair creation and leads to a maximum enhancement of pair production. However, as the field width continues to decrease, the pair creation for combined fields also terminates and the enhancement effect vanishes. Our results are different from those in Fig. 6 of Ref. \cite{Ababekri2019}, where the maximum enhancement factor is less than $2.5$ and the enhancement effect is very limited. This is mainly because the field types we considered are different and the enhancement effect depends on the choice of field parameters. Based on this, in order to further improve pair yield, the field parameters must be optimized.

The reduced momentum spectra for the combined electric fields with $\lambda=3,5,10,30,100/mc$ are shown in Fig. \ref{Fig3}(b). One can see that with the increase of field width the reduced momentum spectrum becomes higher and wider, and its center moves to large momentum. These change fast for $\lambda<10/mc$ and slowly for $\lambda\geq10/mc$. The reason for these changes is roughly consistent with the result of pair creation for only a single strong field with spatial inhomogeneity in \cite{Hebenstreit2011}. However, the momentum spectrum does not show the phenomenon of particle self-bunching with the change of the field width, which also holds true for single strong fields with certain parameters. That is to say, the particle self-bunching phenomenon is not universal in pair production for a spatially inhomogeneous electric field, and it is sensitive to the type of electric fields and field parameters.


\subsubsection{$\lambda_\mathrm{s}>\lambda_\mathrm{w}$}\label{sec:subsec3b2}

\begin{figure}[!ht]
  \centering
  \includegraphics[width=8.5cm]{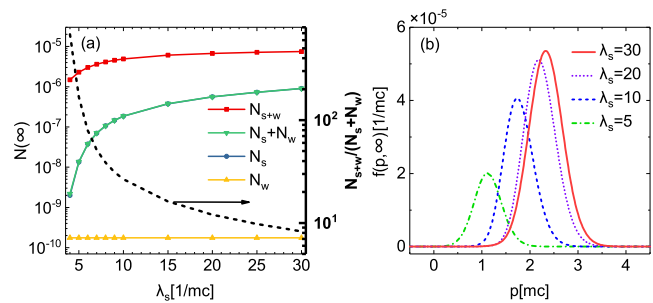}\\
  \caption{(a) Pair yield for combined electric fields $N_\mathrm{s+w}$ (red line and squares), strong fields $N_\mathrm{s}$ (blue line and circles), weak fields $N_\mathrm{w}$ (yellow line and triangles), the sum of the pair yield for strong fields and weak fields $N_\mathrm{s}+N_\mathrm{w}$ (green line and inverted triangles), and the enhancement factor $N_{\mathrm{s+w}}/(N_\mathrm{s}+N_\mathrm{w})$ (black dashed line) as a function of the field width $\lambda_\mathrm{s}$. The right arrow indicates the axis corresponding to the enhancement factor. (b) Momentum spectra for the combined electric fields with $\lambda_\mathrm{s}=5,10,20,30/mc$. Other field parameters are $\lambda_\mathrm{w}=3/mc$, $\omega_\mathrm{s}=0.02mc^2$, $\omega_\mathrm{w}=0.5mc^2$,  $\tau=10/mc^2$, and $\phi=0$.}\label{Fig4}
\end{figure}

In Fig. \ref{Fig4}(a), we plot the pair yield for combined electric fields $N_\mathrm{s+w}$ (red line and squares), strong fields $N_\mathrm{s}$ (blue line and circles), weak fields $N_\mathrm{w}$ (yellow line and triangles), the sum of the pair yield for strong fields and weak fields $N_\mathrm{s}+N_\mathrm{w}$ (green line and inverted triangles), and the enhancement factor $N_{\mathrm{s+w}}/(N_\mathrm{s}+N_\mathrm{w})$ (black dashed line) as a function of the field width $\lambda_\mathrm{s}$. Note that the field width of weak fields $\lambda_\mathrm{w}=3/mc$ is fixed, and the field width of strong fields can change. Moreover, the reduced pair yield and momentum spectrum are not adopted in Fig. \ref{Fig4} as well as in Fig. \ref{Fig5} because they cannot be well defined, but this does not affect the comparison of enhancement factors.


As can be seen from Fig. \ref{Fig4}(a), although the pair yield for strong fields grows linearly with increasing the field width of strong fields for large field width, it is not linear for combined fields and the pair yield increases more and more slowly with the field width, because the dynamical assistance only works in the superposition region of weak fields and strong fields beyond which pair production is severely suppressed. When the field width reaches $30/mc$, the pair yield can increase to $7.6\times10^{-6}$, but it is much less than the pair yield $4\times10^{-5}$ in the case of $\lambda_\mathrm{s}=\lambda_\mathrm{w}=30/mc$. Furthermore, with the increase of the field width of strong fields, the enhancement factor decreases from about $700$ to about $8$, which is also always much smaller than the case that the field widths of weak fields and strong fields are equal. From Fig. \ref{Fig4}(b), it can be seen that the characteristics of the momentum spectrum are similar to those in the case of $\lambda_\mathrm{s}=\lambda_\mathrm{w}$. This is because in both cases the strong low-frequency field plays a dominant role in the pair creation, and the influence of the weak high-frequency field is insignificant. 

\subsubsection{$\lambda_\mathrm{s}<\lambda_\mathrm{w}$}\label{sec:subsec3b3}

\begin{figure}[!ht]
  \centering
  \includegraphics[width=8.5cm]{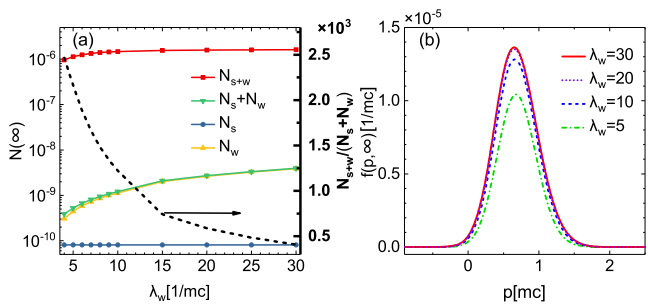}\\
  \caption{Pair yield for combined electric fields $N_\mathrm{s+w}$ (red line and squares), strong fields $N_\mathrm{s}$ (blue line and circles), weak fields $N_\mathrm{w}$ (yellow line and triangles), the sum of the pair yield for strong fields and weak fields $N_\mathrm{s}+N_\mathrm{w}$ (green line and inverted triangles), and the enhancement factor $N_{\mathrm{s+w}}/(N_\mathrm{s}+N_\mathrm{w})$ (black dashed line) as a function of the field width $\lambda_\mathrm{w}$. The right arrow indicates the axis corresponding to the enhancement factor. (b) Momentum spectra for the combined electric fields with $\lambda_\mathrm{w}=5,10,20,30/mc$. Other field parameters are $\lambda_\mathrm{s}=3/mc$, $\omega_\mathrm{s}=0.02mc^2$, $\omega_\mathrm{w}=0.5mc^2$,  $\tau=10/mc^2$, and $\phi=0$.}\label{Fig5}
\end{figure}

When the field width of strong fields $\lambda_\mathrm{s}=3/mc$ is fixed and the field width of strong fields changes, the pair yield for combined electric fields $N_\mathrm{s+w}$ (red line and squares), strong fields $N_\mathrm{s}$ (blue line and circles), weak fields $N_\mathrm{w}$ (yellow line and triangles), the sum of the pair yield for strong fields and weak fields $N_\mathrm{s}+N_\mathrm{w}$ (green line and inverted triangles), and the enhancement factor $N_{\mathrm{s+w}}/(N_\mathrm{s}+N_\mathrm{w})$ (black dashed line) as a function of the field width $\lambda_\mathrm{w}$ are drawn in Fig. \ref{Fig5}(a).


Similar to the result in the case of $\lambda_\mathrm{s}>\lambda_\mathrm{w}$, although the pair yield increases linearly with the increase of the field width of weak fields for large field width, the growth of pair yield for combined fields is not linear and the pair yield remains nearly constant. In this case, the maximum pair yield is $1.4\times10^{-5}$ which is still smaller than  $4\times10^{-5}$ in the case of $\lambda_\mathrm{s}=\lambda_\mathrm{w}=30/mc$, and the enhancement factor decreases from about $2500$ to about $400$ with increasing the field width of weak fields. However, this time the enhancement effect for larger field widths of weak fields is much stronger than that in the case of $\lambda_\mathrm{s}=\lambda_\mathrm{w}$. And because when $\omega_\mathrm{s}=0$ and $\lambda_\mathrm{w}=30/mc$, the combined field can be approximately a superposition field of a strong static field and a spatially uniform but fast oscillating field, it means that the combination of strong static fields and weak high-frequency fields with spatial homogeneous is more effective for enhancing pair production.

The momentum spectrum of created pairs for combined fields in Fig. \ref{Fig5}(b) shows that its maximum value grows continuously with increasing the field width of weak fields. For small field widths the maximum value increases rapidly, while for large field widths the growth slows down to almost zero, which is consistent with the trend of the relation between pair yield and the field width of weak fields. Moreover, the momentum spectrum has only one peak, and its centre does not change with the field width. This can be understood as follows. Particles created from vacuum in the presence of a single weak high-frequency field can be accelerated in the direction of positive and negative $z$-axis and two peaks form in the momentum spectrum. However, when the weak field combines with a strong low-frequency field the pair creation will be dominated by the combined field near its maximum values $E(z=0,t=0)=0.25E_\mathrm{cr}$, and the created particles can only be accelerated in one direction and only one peak forms in the momentum spectrum. Since the weak field is so weak that it has little contribution to the acceleration of created particles especially beyond the superposition region of strong fields and weak fields, the center of momentum spectrum is almost unchanged with the field width. Note that these results are not a universal phenomenon, and they are related to the choice of field parameters.

In the above three cases, all the enhancement factors decrease with the increase of variable field width. When the field width of weak fields is fixed, the change of field width of strong fields has a great effect on the enhancement factor (from $700$ to $8$). When the field width of strong fields is fixed, the change of the field width of weak fields has a little effect on the enhancement factor (from $2500$ to $400$). When the field widths of strong and weak fields change at the same time, the enhancement factor changes between the above two cases. So a strong static field combines with a spatially homogeneous and time varying field can better enhance pair creation process.

\subsection{Effect of the pulse duration on DASPP}\label{sec:subsec3c}

\begin{figure}[!ht]
  \centering
  \includegraphics[width=8.5cm]{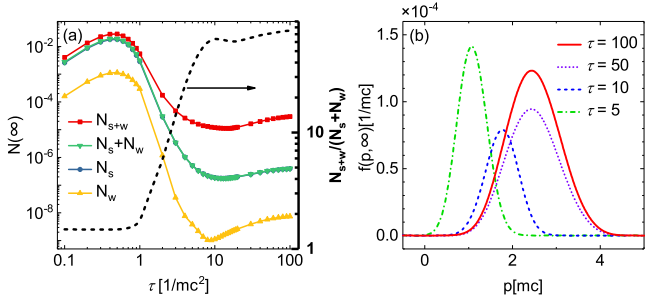}\\
  \caption{Pair yield for combined electric fields $N_\mathrm{s+w}$ (red line and squares), strong fields $N_\mathrm{s}$ (blue line and circles), weak fields $N_\mathrm{w}$ (yellow line and triangles), the sum of the pair yield for strong fields and weak fields $N_\mathrm{s}+N_\mathrm{w}$ (green line and inverted triangles), and the enhancement factor $N_{\mathrm{s+w}}/(N_\mathrm{s}+N_\mathrm{w})$ (black dashed line) as a function of the pulse duration $\tau$. The right arrow indicates the axis corresponding to the enhancement factor. (b) Momentum spectra for the combined electric fields with $\tau=5,10,50,100/mc^2$. Other field parameters are $\lambda_\mathrm{s}=\lambda_\mathrm{w}=10/mc$, $\omega_\mathrm{s}=0.02mc^2$, $\omega_\mathrm{w}=0.5mc^2$, and $\phi=0$.}\label{Fig6}
\end{figure}

To study the effect of pulse durations on DASPP, the pair yield for combined electric fields $N_\mathrm{s+w}$ (red line and squares), strong fields $N_\mathrm{s}$ (blue line and circles), weak fields $N_\mathrm{w}$ (yellow line and triangles), the sum of the pair yield for strong fields and weak fields $N_\mathrm{s}+N_\mathrm{w}$ (green line and inverted triangles), and the enhancement factor $N_{\mathrm{s+w}}/(N_\mathrm{s}+N_\mathrm{w})$ (black dashed line) as a function of the pulse duration $\tau$ are plotted in Fig. \ref{Fig6}(a). The momentum spectra for the combined electric fields with $\tau=5,10,50,100/mc^2$ are shown in Fig. \ref{Fig6}(b).


From Fig. \ref{Fig6}(a), we find that with the increase of pulse duration the pair yield for strong, weak, and combined fields first increases, then decreases and increases again, which means that there is a maximum value and a minimum value. The overall trend of the relation between pair yield and the pulse duration is similar to the result for a spatially homogeneous and time-dependent Sauter pulse shown in \cite{Kohlfurst2013} and the effect of spatial inhomogeneity is not obvious. Specifically, when $\tau\leq5/mc^2$, the strong, weak, and combined fields can be approximately regarded as spatially uniform and time-varying Sauter fields, so the trend of the relation between pair yield and the pulse duration is basically the same as previous results. When $5/mc^2<\tau<30/mc^2$, despite the effect of spatial inhomogeneity, the overall trend is similar to that for Sauter pulses. When $\tau\geq30/mc^2$, the pair yield gradually slows down with the increase of pulse duration and do not meet the linear relationship found in Fig. 5-10 of Ref. \cite{Kohlfurst2015}, which is mainly affected by field profiles. Studying the relation between the enhancement factor and the pulse duration, it is found that when $\tau\leq1/mc^2$ the enhancement factor is less than $2$ and almost does not change with the pulse duration, because pair creation is dominated by the pulse duration at this time, and the dynamically assisted mechanism fails. When $1/mc^2<\tau\leq10/mc^2$, the enhancement factor increases sharply to about $60$ with increasing the pulse duration. When $\tau>10/mc^2$, the trend becomes slow, reaching about $80$ at $\tau=100/mc^2$, and the pair yield is enhanced by nearly two orders of magnitude. This means that the increase of pulse width is more effective to enhance pair production. Moreover, the local minimum near $\tau=18/mc^2$ is caused by the different pulse durations corresponding to the minimum values of pair yield for strong fields and weak fields.

The momentum spectrum in Fig. \ref{Fig6}(b) shows that the trend of the relation between its maximum value and the pulse duration is similar to that between pair yield and the pulse duration, that is, it first increases, then decreases and increases again. In addition, when $\tau\leq30/mc^2$, the momentum spectrum shifts continually to the larger momentum with the increase of the pulse duration. In particular, when $\tau\leq5/mc^2$ the center of momentum spectrum can be determined by $qE_\mathrm{s}\tau$, because the strong field, the weak field and the combined field can be approximately regarded as a spatially homogeneous and time-dependent Sauter pulse. The reason for the above finding is that with the increase of the pulse duration, the duration time of electric fields becomes increasingly long, and the produced particles will gain a longer time to accelerate, which finally makes the momentum spectrum keep shifting to the larger momentum. When $\tau>30/mc^2$, the center and width of the momentum spectrum is almost not change with the increase of pulse duration. This result can be explained as follows. Due to the limitation of the spatial extent of electric fields, the created particles after being accelerated for a period of time will run out of the electric field region, and can no longer be accelerated. Therefore, even if the duration time of electric fields increases continuously, particles will not be accelerated to a larger momentum. Furthermore, for a larger pulse duration, the acceleration of the particles is basically the same from the time when the particles are produced at the maximum field strength to the time when the particles are accelerated out of the electric field region, so the center and the width of momentum spectrum almost does not change with increasing the pulse duration.

\subsection{Effect of the relative phase on DASPP}\label{sec:subsec3d}

\begin{figure}[!ht]
  \centering
  \includegraphics[width=8.5cm]{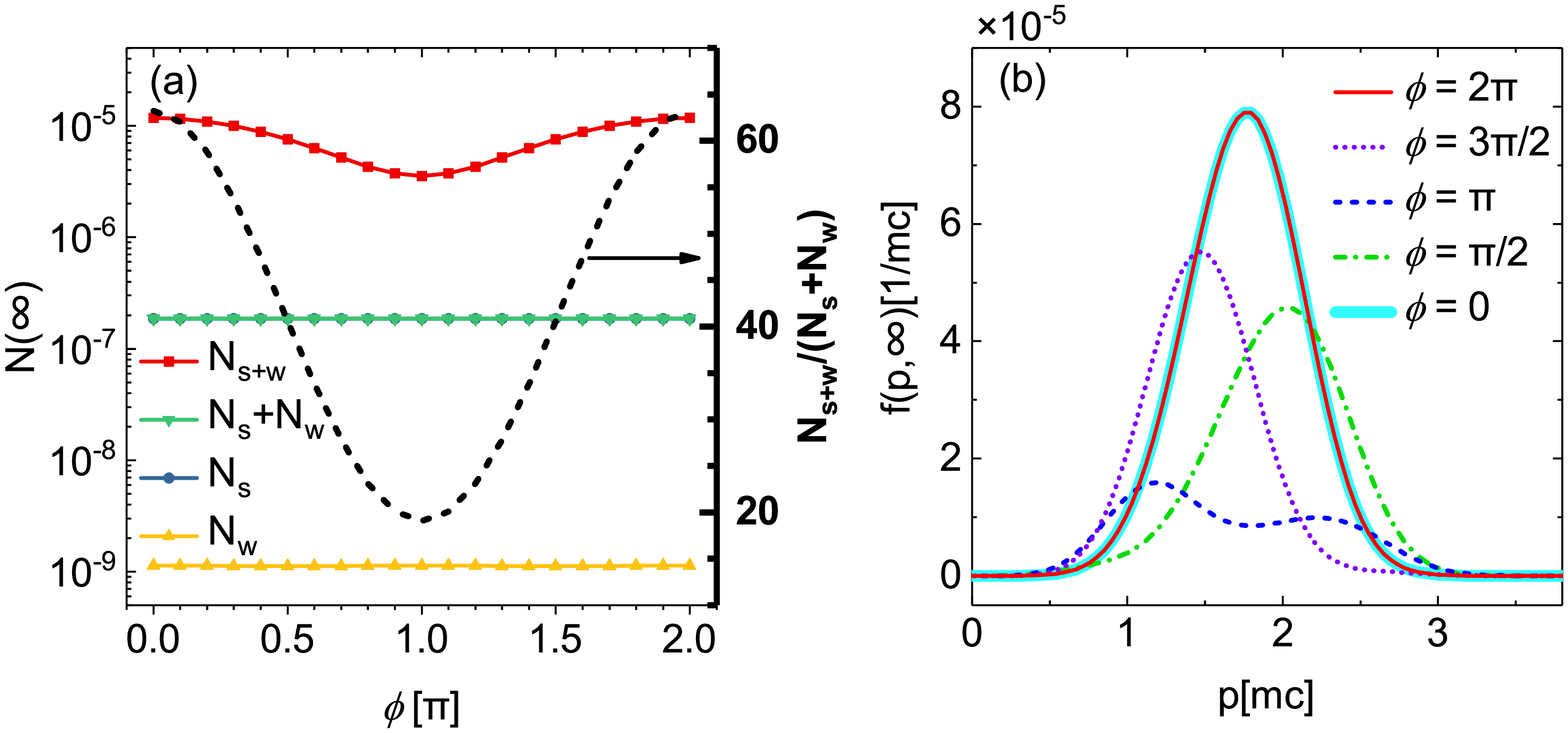}\\
  \caption{Pair yield for combined electric fields $N_\mathrm{s+w}$ (red line and squares), strong fields $N_\mathrm{s}$ (blue line and circles), weak fields $N_\mathrm{w}$ (yellow line and triangles), the sum of the pair yield for strong fields and weak fields $N_\mathrm{s}+N_\mathrm{w}$ (green line and inverted triangles), and the enhancement factor $N_{\mathrm{s+w}}/(N_\mathrm{s}+N_\mathrm{w})$ (black dashed line) as a function of the relative phase $\phi$. The right arrow indicates the axis corresponding to the enhancement factor. (b) Momentum spectra for the combined electric fields with $\phi=0,\pi/2,\pi,3\pi/2,2\pi$. Other field parameters are $\lambda_\mathrm{s}=\lambda_\mathrm{w}=10/mc$, $\omega_\mathrm{s}=0.02mc^2$, $\omega_\mathrm{w}=0.5mc^2$, and $\tau=10/mc^2$.}\label{Fig7}
\end{figure}

The pair yield for combined electric fields $N_\mathrm{s+w}$ (red line and squares), strong fields $N_\mathrm{s}$ (blue line and circles), weak fields $N_\mathrm{w}$ (yellow line and triangles), the sum of the pair yield for strong fields and weak fields $N_\mathrm{s}+N_\mathrm{w}$ (green line and inverted triangles), and the enhancement factor $N_{\mathrm{s+w}}/(N_\mathrm{s}+N_\mathrm{w})$ (black dashed line) as a function of the relative phase $\phi$ are given to explore the effect of the relative phase on DASPP. Note that $\phi$ is the relative carrier envelope phase and the initial phase of the strong field is set to $0$, so the pair yield does not change with $\phi$ for the strong field. For a single weak field, however, the pair yield oscillates with $\phi$ with the period $\pi$, but this change is so small that it can hardly be seen from Fig. \ref{Fig7}(a).


For combined fields, it is found that the pair yield is greatly increased and changes with the relative phase with the period $2\pi$. It achieves the minimum value $3.6\times10^{-6}$ at $\phi=\pi$ and the maximum value $1.2\times10^{-5}$ at $\phi=0$ or $2\pi$. Moreover, the enhancement factor also varies with the relative phase with the period $2\pi$. Its minimum value is about $20$ achieved at $\phi=\pi$ and the maximum value is about $63$ achieved at $\phi=0$ or $2\pi$. This indicates that the relative phase between strong fields and weak fields has an obvious enhancement effect on the pair creation process. Here it is easy to understand that the pair yield and enhancement factor have minimum values at $\phi=\pi$, because at this time, when the strong fields and weak fields reach their maximum values respectively, their directions are opposite and the maximum field strength of the combined field becomes $0.2-0.05=0.15$, which makes the pair creation process as well as its enhancement become weaker than those for other values of $\phi$.

The momentum spectra for combined electric fields with $\phi=0,\pi/2,\pi,3\pi/2,2\pi$ are plotted in Fig. \ref{Fig7}(b). It can be seen that the cycle of the pair creation process varying with the relative phase is $2\pi$ rather than $\pi$, because the momentum spectra for $\phi=\pi/2$ and $\phi=3\pi/2$ are different from each other, while for $\phi=0$ and $\phi=2\pi$ they are the same. Moreover, we also find that the center of the momentum spectrum is on the right for $\phi=\pi/2$ and on the left for $\phi=3\pi/2$ comparing with that for $\phi=0$, and the momentum spectrum has two peaks for $\phi=\pi$. To explain this result, we must start from the configuration of the combined field. When $\phi=\pi/2$, the maximum value of the combined field appears at $t<0$, so the created particles will be accelerated in the electric field for a long time. However, when $\phi=3\pi/2$, the maximum value of the combined field appears at $t>0$, the created particles can only be accelerated for a short time. This eventually leads to the deviation of the center of momentum spectrum in different directions. In the case of $\phi=\pi$, the maximum values of the combined field are symmetrically distributed at $t<0$ and $t>0$, and the particles produced at the maximum value of $t<0$ can be accelerated much longer than that at $t>0$, which forms two peaks in the momentum spectrum, one on the right and the other on the left. The reason why the peak value on the right is smaller than that on the left is mainly due to the influence of the spatial inhomogeneity of electric fields, which makes the right peak wider and lower.

\section{Conclusions and Discussions}\label{sec:sec4}

In summary, the equivalence between the CQFT and the QKT for pair creation in arbitrary spatially uniform and time-varying electric fields is proved theoretically first, and then it is generalized numerically to one-dimensional time-dependent electric fields. Finally, the effect of field parameters on the dynamically assisted Schwinger pair production in inhomogeneous two-color electric fields is investigated by employing the CQFT approach.

The study of the effect of field frequencies on the DASPP shows that the pair production can be enhanced three orders of magnitude for appropriate field frequency, but sometimes it can also be suppressed. For the effect of field widths, the enhancement of pair creation increases with the decrease of variable field width and can be improved three orders of magnitude as well for small field widths. When the field width of strong fields is less than that of weak fields, the enhancement of pair creation has a relatively small change with increasing the field width and it can still be improved two orders of magnitude for large field widths. Note that when the field width becomes very small the enhancement factor varying with the field width may have one or two maximum values for some field parameters due to the existence of threshold widths of strong fields and weak fields. For the effect of pulse durations, the enhancement of pair creation grows with the increase of pulse duration and can be improved nearly two orders of magnitude. For the effect of relative phases, the enhancement of pair creation increases with the pulse duration and can also be improved nearly two orders of magnitude, which indicates that the relative phase is also a significant field parameter to affect the pair creation process. These results indicate that the DASPP still holds true for spatially inhomogeneous electric fields and provide abundant information for the forthcoming optimal theory which can globally maximize the pair yield in spatially inhomogeneous electric fields within certain range of field parameters.

The study of the effect of the field width, the pulse duration, and the relative phase on momentum spectrum of created pairs further deepens our understanding of important characteristics of pair creation in combined fields. Analyzing the momentum spectrum, it is also found that there is no obvious multiphoton peak structure in it. This is mainly due to the limitation of the field strength and the pulse duration of weak fields. On the one hand, the field strength of weak high-frequency fields is so small that the pair yield produced by absorbing a larger number of photons is very low, which is difficult to be reflected. On the other hand, since the presence of multiphoton peaks in the momentum spectrum needs enough number of oscillation periods in the electric field envelope, and for a given field frequency the number of cycles reduces with decreasing the pulse duration, multiphoton peaks will not appear for a very small pulse duration. Therefore, the exploration of the momentum spectrum with multiphoton peaks in DASPP will be a research focus in our future work.

In addition, due to the fact that the equivalence between the CQFT and the QKT for pair creation in time-dependent electric fields and one-dimensional time-varying electric fields in spinor QED has been proved, and can be easily generalized to the scalar QED, the generalization of this equivalence to high-dimensional time-varying electric fields with or without magnetic fields is a very feasible work and will also be studied in future.

\acknowledgments
The work is supported by the National Natural Science Foundation of China (NSFC) under Grants  No. 11705278 and No. 11974419, in part by the National Key R\&D Program of China under Grant No. 2018YFA0404802, and by the Fundamental Research Funds for the Central Universities.

\appendix

\section{Quantum Vlasov equation}\label{app:Appa}

For a spatially homogeneous and time-dependent electric field with temporal gauge $A_\mu(\mathbf{x},t)=(0,-\mathbf{A}(t)/c)=(0,0,0,-A_z(t)/c)$ and $E_z(t)=-dA_z(t)/dt$, equation (\ref{eqn:DiracEquation1}) is reduced to
\begin{equation}\label{eqn:DiracEquation10}
\big\{i\gamma^{0}\partial_t+i\bm{\gamma}\cdot[c\bm{\nabla}-i q \mathbf{A}(t)]-m c^2\big\}\Psi(\mathbf{x},t)=0.
\end{equation}
Furthermore, the Dirac field can be expanded into Fourier modes as
\begin{equation}\label{eqn:FourierDeco0}
\Psi(\mathbf{x},t)=\int\frac{d^3k}{(2\pi)^3}\Psi_\mathbf{k}(t)e^{i\mathbf{k}\cdot\mathbf{x}},
\end{equation}
where $\mathbf{k}$ is the canonical momentum and the Fourier modes $\Psi_\mathbf{k}(t)$ satisfy
\begin{equation}\label{eqn:DiracEquation11}
\big[i\gamma^{0}\partial_t-\bm{\gamma}\cdot c\mathbf{p}(t)-m c^2\big]\Psi_\mathbf{k}(t)=0,
\end{equation}
where $\mathbf{p}(t)=\mathbf{k}-q\mathbf{A}(t)/c$ is the kinetic momentum. Supposing
\begin{equation}\label{eqn:DiracEquation12}
\Psi_\mathbf{k}(t)=\big[i\gamma^{0}\partial_t-\bm{\gamma}\cdot c\mathbf{p}(t)+m c^2\big]\psi_\mathbf{k}(t),
\end{equation}
and then substituting it into Eq. (\ref{eqn:DiracEquation11}), we have
\begin{equation}\label{eqn:PDE1}
\Big[\partial^2_t+\omega_\mathbf{k}^2(t)+iqE_z(t)\gamma^0\gamma^3\Big]\psi_\mathbf{k}(t)=0,
\end{equation}
where $\omega_\mathbf{k}(t)=[(c\mathbf{k}-q\mathbf{A}(t))^2+m^2c^4]^{1/2}$ is the total energy of particles.

Expanding $\psi_\mathbf{k}(t)$ in terms of the eigenvectors of $\gamma^0\gamma^3$ as
\begin{equation}
\psi_\mathbf{k}(t)=\sum_{s=1}^4\chi_\mathbf{k}^s(t)R_s,
\end{equation}
where $R_1\!=\!\frac{1}{\sqrt{2}}\left(
                                   \begin{array}{cccc}
                                     1 & 0 & 1 & 0 \\
                                   \end{array}
                                 \right)^\textsf{T}$,
$R_2\!=\!\frac{1}{\sqrt{2}}\left(
                                   \begin{array}{cccc}
                                     0 & 1 & 0 & -1 \\
                                   \end{array}
                                 \right)^\textsf{T}$,
$R_3\!=\!\frac{1}{\sqrt{2}}\left(
                                   \begin{array}{cccc}
                                     1 & 0 & -1 & 0 \\
                                   \end{array}
                                 \right)^\textsf{T}$, and
$R_4\!=\!\frac{1}{\sqrt{2}}\left(
                                   \begin{array}{cccc}
                                     0 & 1 & 0 & 1 \\
                                   \end{array}
                                 \right)^\textsf{T}$
are the eigenvectors and satisfy $\gamma^0\gamma^3R_{s=\{1,2\}}=+1R_{s=\{1,2\}}$,
$\gamma^0\gamma^3R_{s=\{3,4\}}=-1R_{s=\{3,4\}}$ and $R_r^\dagger R_s=\delta_{rs}$, and inserting the above equation into Eq. (\ref{eqn:PDE1}), we obtain
\begin{equation}
\begin{split}
\big[\partial^2_t+\omega^2_\mathbf{k}(t)+iqE_z(t)\big]\chi^{s=\{1,2\}}_\mathbf{k}(t)=0,\\
\big[\partial^2_t+\omega^2_\mathbf{k}(t)-iqE_z(t)\big]\chi^{s=\{3,4\}}_\mathbf{k}(t)=0.
\end{split}
\end{equation}
These four equations are overdetermined and can be removed by choosing $s=\{1,2\}$ or $s=\{3,4\}$. Here we use the former one and redefine $\chi^{+}_\mathbf{k}(t)=\chi^{1}_\mathbf{k}(t)$ and $\chi^{-}_\mathbf{k}(t)=\chi^{2}_\mathbf{k}(t)$.

Then the field operator at time $t$ can be expressed as
\begin{equation}\label{eqn:FieldOperator1}
\Psi(\mathbf{x},t)=\!\int\!\frac{d^3k}{(2\pi)^3}\sum_{s=1}^2\big[b_{\mathbf{k},s}u_{\mathbf{k},s}(t)
+d_{-\mathbf{k},s}^{\dagger}v_{-\mathbf{k},s}(t)\big]e^{i\mathbf{k}\cdot\mathbf{x}},
\end{equation}
by virtue of the time-independent annihilation operator of particles $b_{\mathbf{k},s}$ and creation operator of antiparticles $d_{-\mathbf{k},s}^{\dagger}$, which fulfil the standard fermionic anticommutation relations, where
\begin{equation}\label{eqn:FieldOperator2}
\begin{split}
u_{\mathbf{k},s}(t)&=\big[i\gamma^{0}\partial_t-\bm{\gamma}\cdot c\mathbf{p}(t)+m c^2\big]\chi^{+}_\mathbf{k}(t)R_s,\\
v_{-\mathbf{k},s}(t)&=\big[i\gamma^{0}\partial_t-\bm{\gamma}\cdot c\mathbf{p}(t)+m c^2\big]\chi^{-}_\mathbf{k}(t)R_s,
\end{split}
\end{equation}
are positive and negative energy states, respectively. Note that Eqs. (\ref{eqn:FieldOperator1}) and (\ref{eqn:FieldOperator2}) can be obtained by the time evolution of the force-free positive and negative energy states
\begin{equation}\label{eqn:FreeState1}
\begin{split}
u_{\mathbf{k},s}(t_0)&=\big[\gamma^{0}\omega_\mathbf{k}-\bm{\gamma}\cdot c\mathbf{k}+m c^2\big]e^{-i\omega_\mathbf{k}t_0}R_s,\quad\\
v_{-\mathbf{k},s}(t_0)&=\big[-\gamma^{0}\omega_\mathbf{k}-\bm{\gamma}\cdot c\mathbf{k}+m c^2\big]e^{+i\omega_\mathbf{k}t_0}R_s,
\end{split}
\end{equation}
according to the Dirac equation for spatially homogeneous electric fields Eq. (\ref{eqn:DiracEquation11}).

Calculating the Hamiltonian from the energy-momentum tensor, it is found that the Hamiltonian in the presence of an electric field is off-diagonal, which indicates that the interpretation of particles and antiparticles is not feasible because of the mixture of positive and negative energy modes. In order to give a meaningful interpretation of particles and antiparticles, one can diagonalize the Hamiltonian by a time-dependent Bogoliubov transformation:
\begin{equation}\label{eqn:BogoliubovTransformation}
\begin{split}
\widetilde{b}_{\mathbf{k},s}(t)&=\alpha_\mathbf{k}(t)b_{\mathbf{k},s}-\beta^*_\mathbf{k}(t)d^\dagger_{-\mathbf{k},s},\\
\widetilde{d}^{\,\dagger}_{-\mathbf{k},s}(t)&=\beta_\mathbf{k}(t)b_{\mathbf{k},s}
+\alpha^*_\mathbf{k}(t)d^\dagger_{-\mathbf{k},s},
\end{split}
\end{equation}
with the only nonzero anticommutators $\{\widetilde{b}_{\mathbf{k},r}(t),\widetilde{b}^{\,\dagger}_{\mathbf{k'},s}(t)\}
=\{\widetilde{d}_{\mathbf{k},r}(t),\widetilde{d}^{\,\dagger}_{\mathbf{k'},s}(t)\}
=(2\pi)^3\delta_{rs}\delta(\mathbf{k}-\mathbf{k'})$ and $|\alpha_\mathbf{k}(t)|^2+|\beta_\mathbf{k}(t)|^2=1$. This transformation is equivalent to expanding the field operator in the adiabatic basis, $\widetilde{u}_{\mathbf{k},s}(t)$ and $\widetilde{v}_{-\mathbf{k},s}(t)$, i.e.,
\begin{equation}\label{eqn:FieldOperator3}
\begin{split}
\Psi(\mathbf{x},t)=\!\!\int\!\frac{d^3k}{(2\pi)^3}\sum_{s=1}^2\big[&\,\widetilde{b}_{\mathbf{k},s}(t)
\widetilde{u}_{\mathbf{k},s}(t)\\
&+\widetilde{d}_{-\mathbf{k},s}^{\,\dagger}(t)\widetilde{v}_{-\mathbf{k},s}(t)\big]
e^{i\mathbf{k}\cdot\mathbf{x}},
\end{split}
\end{equation}
where
\begin{equation}\label{eqn:FieldOperator4}
\begin{split}
\widetilde{u}_{\mathbf{k},s}(t)&=\big[\gamma^{0}\omega_\mathbf{k}(t)-\bm{\gamma}\cdot c\mathbf{p}(t)+m c^2\big]\widetilde{\chi}^{+}_\mathbf{k}(t)R_s,\\
\widetilde{v}_{-\mathbf{k},s}(t)&=\big[-\gamma^{0}\omega_\mathbf{k}(t)-\bm{\gamma}\cdot c\mathbf{p}(t)+m c^2\big]\widetilde{\chi}^{-}_\mathbf{k}(t)R_s,
\end{split}
\end{equation}
with the adiabatic mode functions
\begin{equation}
\widetilde{\chi}^{\pm}_\mathbf{k}(t)=\frac{e^{\mp i\Theta_\mathbf{k}(t_0,t)}}{\sqrt{2\omega_\mathbf{k}(t)[\omega_\mathbf{k}(t)\mp c p_z(t)]}},
\end{equation}
the dynamical phase $\Theta_\mathbf{k}(t_0,t)=\int^t_{t_0}d\tau\,\omega_\mathbf{k}(\tau)$, and $\widetilde{u}_{\mathbf{k},r}^{\,\dagger}(t)\widetilde{u}_{\mathbf{k},s}(t)
=\widetilde{v}_{-\mathbf{k},r}^{\,\dagger}(t)\widetilde{v}_{-\mathbf{k},s}(t)
=\delta_{rs}$, $\widetilde{u}_{\mathbf{k},r}^{\,\dagger}(t)\widetilde{v}_{-\mathbf{k},s}(t)
=\widetilde{v}_{-\mathbf{k},r}^{\,\dagger}(t)\widetilde{u}_{\mathbf{k},s}(t)=0$.
For convenience, we also show the matrix form of the adiabatic basis:
\begin{equation}\label{eqn:FieldOperator40}
\begin{split}
\widetilde{u}_{\mathbf{k},1}(t)&=\frac{1}{\sqrt{2}}\left(
  \begin{array}{c}
   \omega_\mathbf{k}(t)+mc^2-c p_z(t) \\
       -c(p_x+ip_y) \\
       -\omega_\mathbf{k}(t)+mc^2+c p_z(t) \\
       c(p_x+ip_y) \\
  \end{array}
\right)\widetilde{\chi}^{+}_\mathbf{k}(t), \quad\\
\widetilde{u}_{\mathbf{k},2}(t)&=\frac{1}{\sqrt{2}}\left(
  \begin{array}{c}
    c(p_x-ip_y) \\
       \omega_\mathbf{k}(t)+mc^2-c p_z(t) \\
       c(p_x-ip_y) \\
       \omega_\mathbf{k}(t)-mc^2-c p_z(t) \\
  \end{array}
\right)\widetilde{\chi}^{+}_\mathbf{k}(t),\\
\widetilde{v}_{-\mathbf{k},1}(t)&=\frac{1}{\sqrt{2}}\left(
  \begin{array}{c}
    -\omega_\mathbf{k}(t)+mc^2-c p_z(t) \\
       -c(p_x+ip_y) \\
       \omega_\mathbf{k}(t)+mc^2+c p_z(t) \\
       c(p_x+ip_y) \\
  \end{array}
\right)\widetilde{\chi}^{-}_\mathbf{k}(t),\;\,\\
\widetilde{v}_{-\mathbf{k},2}(t)&=\frac{1}{\sqrt{2}}\left(
  \begin{array}{c}
    c(p_x-ip_y) \\
       -\omega_\mathbf{k}(t)+mc^2-c p_z(t) \\
       c(p_x-ip_y) \\
       -\omega_\mathbf{k}(t)-mc^2-c p_z(t) \\
  \end{array}
\right)\widetilde{\chi}^{-}_\mathbf{k}(t).
\end{split}
\end{equation}
Therefore, the particle number density of created pairs for a given canonical momentum $\mathbf{k}$ can be defined as
\begin{equation}\label{eqn:NumberDensity}
F_{\mathbf{k}}(t)=\lim_{V\rightarrow\infty}\sum_{s=1}^2\frac{\langle \mathrm{vac}|\widetilde{b}^{\,\dagger}_{\mathbf{k},s}(t)\widetilde{b}_{\mathbf{k},s}(t)|\mathrm{vac}\rangle}{V}
=2|\beta_\mathbf{k}(t)|^2,
\end{equation}
which is also the definition of the one-particle distribution function. Note that the simply sum over both spin-indices is based on the fact that there is no spin preference in the system with vanishing magnetic fields.

The next thing is to determine the Bogoliubov transformation coefficient $\beta_\mathbf{k}(t)$. Inserting Eq. (\ref{eqn:BogoliubovTransformation}) into Eq. (\ref{eqn:FieldOperator3}) and comparing with Eq. (\ref{eqn:FieldOperator1}), we obtain
\begin{equation}\label{eqn:Relation}
\begin{split} u_{\mathbf{k},s}(t)&=\alpha_{\mathbf{k}}(t)\widetilde{u}_{\mathbf{k},s}(t)
+\beta_{\mathbf{k}}(t)\widetilde{v}_{-\mathbf{k},s}(t), \\ v_{-\mathbf{k},s}(t)&=-\beta^*_{\mathbf{k}}(t)\widetilde{u}_{\mathbf{k},s}(t)
+\alpha^*_{\mathbf{k}}(t)\widetilde{v}_{-\mathbf{k},s}(t).
\end{split}
\end{equation}
Substituting Eqs. (\ref{eqn:FieldOperator2}) and (\ref{eqn:FieldOperator40}) into the above equations, it finds
\begin{equation}\label{eqn:BogoliubovCoefficient}
\begin{split}
\alpha_\mathbf{k}(t)&=\widetilde{u}^{\,\dagger}_{\mathbf{k},s}(t)u_{\mathbf{k},s}(t)=
[\widetilde{v}^{\,\dagger}_{-\mathbf{k},s}(t)v_{-\mathbf{k},s}(t)]^* \\
&=i\epsilon_\perp\widetilde{\chi}^{-}_\mathbf{k}(t)[\partial_t
-i\omega_\mathbf{k}(t)]\chi^{+}_\mathbf{k}(t),\\
\beta_\mathbf{k}(t)&=\widetilde{v}^{\,\dagger}_{-\mathbf{k},s}(t)u_{\mathbf{k},s}(t)=
-[\widetilde{u}^{\,\dagger}_{\mathbf{k},s}(t)v_{-\mathbf{k},s}(t)]^*\\
&=-i\epsilon_\perp\widetilde{\chi}^{+}_\mathbf{k}(t)[\partial_t
+i\omega_\mathbf{k}(t)]\chi^{+}_\mathbf{k}(t),
\end{split}
\end{equation}
where $s$ denotes a specific spin here and $\epsilon_\perp=(c^2\mathbf{k}_\perp^2+m^2c^4)^{1/2}=[c^2(k_x^2+k_y^2)+m^2c^4]^{1/2}$ is the transverse energy.
Then the time derivatives of the Bogoliubov coefficients are
\begin{equation}\label{eqn:TimeDerivative}
\begin{split}
\dot{\alpha}_\mathbf{k}(t)&=\frac{1}{2}W_\mathbf{k}(t)\beta_\mathbf{k}(t)e^{2i\Theta_\mathbf{k}(t_0,t)},\\
\dot{\beta}_\mathbf{k}(t)&=-\frac{1}{2}W_\mathbf{k}(t)\alpha_\mathbf{k}(t)e^{-2i\Theta_\mathbf{k}(t_0,t)}.
\end{split}
\end{equation}
where dot denotes the total time derivative and   $W_\mathbf{k}(t)=cqE_z(t)\epsilon_\perp/2\omega^2_\mathbf{k}(t)$.

With the help of the adiabatic particle correlation function
\begin{equation}
\begin{split}
C_{\mathbf{k}}(t)&=\lim_{V\rightarrow\infty}\sum_{s=1}^2\frac{\langle \mathrm{vac}|\widetilde{d}^{\,\dagger}_{-\mathbf{k},s}(t)\widetilde{b}^{\,\dagger}_{\mathbf{k},s}(t)|\mathrm{vac}\rangle}{V}\\
&=2\alpha^*_\mathbf{k}(t)\beta_\mathbf{k}(t),
\end{split}
\end{equation}
one can derive the quantum Vlasov equation (QVE) in integral-differential form:
\begin{equation}\label{eqn:QKT}
\dot{F}_\mathbf{k}(t)=W_\mathbf{k}(t)\int_{t_0}^t\!dt'W_\mathbf{k}(t')
[1-F_\mathbf{k}(t')]\cos[2\Theta_\mathbf{k}(t',t)],
\end{equation}
with $\Theta_\mathbf{k}(t',t)=\int^t_{t'}d\tau\,\omega_\mathbf{k}(\tau)$ and the vacuum initial condition $F_\mathbf{k}(-\infty)=0$.

The total number density of created pairs is
\begin{equation}\label{eqn:NumberDensity1}
\mathcal{N}(t)=\int\frac{d^3k}{(2\pi)^3}F_\mathbf{k}(t).
\end{equation}

From the derivation of QVE, it is found that any complete basis of spinors that can diagonalize the Hamiltonian is able to give a definition of particles, which leads to that the definition of particles during the existence of external electric fields is not unique and has no clear physical meaning. So all of the particles defined during the interaction time are interpreted as quasi-particles which are a mixture of real particles and virtual particles. However, no matter how to define the quasi-particles, all of them will become real particles and give the same and definite number density of created pairs when the electric fields vanishes at $t\rightarrow+\infty$ \cite{Unger2019,Dabrowski2014}. Therefore, the application of QVE is not limited by adiabatic conditions though the adiabatic basis is used in its derivation. However, if the instantaneous energy of particles $\omega_\mathbf{k}(t)$ varies slowly, the definition in Eq. (\ref{eqn:NumberDensity}) can have an interpretation of adiabatic particle number during the interaction time, otherwise it should be interpreted as quasi-particle number.

\section{DHW formalism in QED$_{1+1}$}\label{app:Appb}

In this appendix, we briefly derive the equal-time DHW formalism in QED$_{1+1}$ from that in QED$_{3+1}$.

For a one-dimensional time-dependent electromagnetic field with temporal gauge $A_\mu(\mathbf{x},t)=(0,-\mathbf{A}(\mathbf{x},t)/c)$, the electric and magnetic fields are \begin{equation}
  \mathbf{E}(\mathbf{x},t)=-\frac{1}{c}\frac{\partial}{\partial t}\mathbf{A}(\mathbf{x},t), \quad\mathbf{B}(\mathbf{x},t)=\bm{\nabla}\times\mathbf{A}(\mathbf{x},t).
\end{equation}
The Wigner function is defined as the vacuum expectation value of the Wigner operator which is the Fourier transform of the equal-time commutator of two Dirac field operators in the Heisenberg picture $\mathcal{C}(\mathbf{x},\mathbf{r},t)$ with respect to the relative coordinate
$\mathbf{r}$:
\begin{eqnarray}\label{eqn:WignerFunction}
\mathcal{W}(\mathbf{x},\mathbf{p},t)=\int &&\hspace{-0.4cm}d^3r\,e^{-i\mathbf{p}\cdot\mathbf{r}}\mathcal{C}(\mathbf{x},\mathbf{r},t)\,\nonumber\\
=\int &&\hspace{-0.4cm}d^3r\,e^{-i\mathbf{p}\cdot\mathbf{r}}e^{-i\frac{q}{c}\int_{-1/2}^{1/2}d\xi\mathbf{A}(\mathbf{x}
+\xi\mathbf{r},t)\cdot\mathbf{r}\,}\,\\
&&\hspace{-0.4cm}\times\langle\mathrm{vac}|\tfrac{1}{2}[\bar{\Psi}(\mathbf{x}-\tfrac{\mathbf{r}}{2},t),
\Psi(\mathbf{x}+\tfrac{\mathbf{r}}{2},t)]|\mathrm{vac}\rangle. \nonumber
\end{eqnarray}
where $\mathbf{x}$ is the center-of-mass coordinate. The second exponential function in the integrand on the right hand side of Eq. (\ref{eqn:WignerFunction}) is called Wilson line factor which is introduced to keep gauge invariance. Furthermore, a straight line is chosen as the integration path in this factor to introduce a well defined kinetic momentum $\mathbf{p}$. Note that a Hartree approximation is used here, so the electromagnetic field is treated as a C-number field instead of a Q-number one. 

Taking the time derivative of Eq.~(\ref{eqn:WignerFunction}) and applying the Dirac equation Eq. (\ref{eqn:DiracEquation10}) with $\mathbf{A}(t)$ replaced by $\mathbf{A}(\mathbf{x},t)$, we can obtain the equation of motion for the Wigner function:
\begin{equation}\label{eqn:EOM}
  D_t\mathcal{W}=-\frac{c}{2}\mathbf{D}\cdot\left[\gamma^0\bm{\gamma},\mathcal{W}\right]
  \!-ic\mathbf{\Pi}\cdot\left\{\gamma^0\bm{\gamma},\mathcal{W}\right\}
  \!-imc^2\left[\gamma^0,\mathcal{W}\right]
  ,
\end{equation}
where $D_t$, $\mathbf{D}$ and $\mathbf{\Pi}$ denote the pseudo-differential operators
\begin{alignat}{6}
  &D_t&\ =\ &\ \ \partial_t \ &\ + \ & q  \int_{-1/2}^{1/2}{d\xi\,\mathbf{E}(\mathbf{x}+i\xi\partial_\mathbf{p}},t)\cdot\partial_\mathbf{p} \ , \nonumber
  \\
  \label{eqn:DifferentialOperator}
  &\mathbf{D}&\ =\ &\ \bm{\nabla} \ & + \ & q  \int_{-1/2}^{1/2}{d\xi\,\mathbf{B}(\mathbf{x}+i\xi\partial_\mathbf{p},t)
  \times\partial_\mathbf{p}} \ ,
  \\
  &\mathbf{\Pi}&\ =\ &\ \ \mathbf{p}\ &\ - \ & iq  \int_{-1/2}^{1/2}{d\xi\,\xi\,\mathbf{B}(\mathbf{x}+i\xi\partial_\mathbf{p},t)
  \times\partial_\mathbf{p}} \ . \nonumber
\end{alignat}

The Wigner function $\mathcal{W}(\mathbf{x},\mathbf{p},t)$ can be expanded in terms of a complete basis set
$\{\mathbbm{1},\gamma_5,\gamma^\mu,\gamma^\mu\gamma_5,\sigma^{\mu\nu}
=:\frac{i}{2}[\gamma^\mu,\gamma^\nu]\}$ and $16$ irreducible components (DHW functions), scalar
$\mathbbm{s}(\mathbf{x},\mathbf{p},t)$, pseudoscalar
$\mathbbm{p}(\mathbf{x},\mathbf{p},t)$, vector $\mathbbm{v}_\mu(\mathbf{x},\mathbf{p},t)$,
axialvector $\mathbbm{a}_\mu(\mathbf{x},\mathbf{p},t)$ and tensor
$\mathbbm{t}_{\mu\nu}(\mathbf{x},\mathbf{p},t)$ as
\begin{equation}\label{eqn:Expand}
  \mathcal{W}(\mathbf{x},\mathbf{p},t)=\frac{1}{4}\left(\mathbbm{1}\mathbbm{s}+i\gamma_5\mathbbm{p}
+\gamma^\mu\mathbbm{v}_\mu+\gamma^\mu\gamma_5\mathbbm{a}_\mu+\sigma^{\mu\nu}\mathbbm{t}_{\mu\nu}\right).
  \
\end{equation}
Substituting the above decomposition into Eq.~(\ref{eqn:EOM}), we obtain a partial differential equation system for the 16 DHW functions as
\begin{alignat}{8}
  \label{eqn:DHW1}
  &D_t\,\mathbbm{s}\ &\  &\  &\ -&\ 2c\mathbf{\Pi}\cdot\bm{\mathbbm{t}_{1}}\ &\ =\ &\ &\ & 0\\
  \label{eqn:DHW2}
  &D_t\,\mathbbm{p}\ &\  &\  &\ +&\ 2c\mathbf{\Pi}\cdot\bm{\mathbbm{t}_{2}}\ &\ =\ &\ &2mc^2\ & \mathbbm{a}_0\\
  \label{eqn:DHW3}
  &D_t\,\mathbbm{v}_0\ &\ +&\ c\mathbf{D}\cdot\bm{\mathbbm{v}}\ &\  &\  &\ =\ &\ &\ & 0\\
  \label{eqn:DHW4}
  &D_t\,\mathbbm{a}_0\ &\ +&\ c\mathbf{D}\cdot\bm{\mathbbm{a}}\ &\  &\  &\ =\ &\ &2mc^2\ & \mathbbm{p}\\
  \label{eqn:DHW5}
  &D_t\,\bm{\mathbbm{v}}\ &\ + &\ c\mathbf{D}\,\mathbbm{v}_0\ &\ +&\ 2c\mathbf{\Pi}\times\bm{\mathbbm{a}}\ &\ =\ &\ -&2mc^2\ & \bm{\mathbbm{t}_1} \\
  \label{eqn:DHW6}
  &D_t\,\bm{\mathbbm{a}}\ &\ +&\ c\mathbf{D}\,\mathbbm{a}_0\ &\ +&\ 2c\mathbf{\Pi}\times\bm{\mathbbm{v}}\ &\ =\ &\ &\ & \mathbf{0}\\
  \label{eqn:DHW7}
  &D_t\,\bm{\mathbbm{t}_{1}}\ &\ +&\ c\mathbf{D}\times\bm{\mathbbm{t}_2}\ & + &\ 2c\mathbf{\Pi}\,\mathbbm{s}&\ =\ &\ &2mc^2\ & \bm{\mathbbm{v}}\\
  \label{eqn:DHW8}
  &D_t\,\bm{\mathbbm{t}_{2}}\ &\ -&\ c\mathbf{D}\times\bm{\mathbbm{t}_1}\ & - &\ 2c\mathbf{\Pi}\,\mathbbm{p}&\ =\ &\ & &\mathbf{0}
\end{alignat}
with
\begin{equation}
  \left(\bm{\mathbbm{t}_1}\right)^i=2\mathbbm{t}^{i0}=2\mathbbm{t}_{0i}, \qquad \left(\bm{\mathbbm{t}_2}\right)^i=\epsilon^{ijk}\mathbbm{t}_{jk}.
\end{equation}

For a $1+1$ dimensional electromagnetic field with vanishing magnetic field $A_\mu(\mathbf{x},t)=(0,0,0,-A_z(z,t)/c)$ where $E_z(z,t)=-\partial A_z(z,t)/(c \partial t)$, there are only two $4\times4$ Dirac gamma matrices, $\gamma^0$ and $\gamma^1$. Thus, the decomposition Eq. (\ref{eqn:Expand}) is reduced to
\begin{equation}\label{eqn:Expand1}
  \mathcal{W}(z,p_z,t)=\frac{1}{4}(\mathbbm{1}\mathbbm{s}
+\gamma^0\mathbbm{v}_0+\gamma^1\mathbbm{v}_z+i\gamma^0\gamma^1\mathbbm{t}_{1z}).
\end{equation}
Correspondingly, the DHW functions $\mathbbm{p}$, $\mathbbm{a}_0$, $\bm{\mathbbm{a}}$ and $\bm{\mathbbm{t}_{2}}$ in Eqs. (\ref{eqn:DHW1})-(\ref{eqn:DHW8}) vanish. So we obtain the DHW formalism in QED$_{1+1}$:
\begin{alignat}{8}
  \label{eqn:DHW11}
  &D_t\,\mathbbm{s}\ &\  &\  &\ -&\ 2cp_z\,\mathbbm{t}_{1z}\ &\ =\ &\ &\ & 0 &\ ,  \\
  &D_t\,\mathbbm{v}_0\ &\ +&\ c\partial_z\mathbbm{v}_z\ &\  &\  &\ =\ &\ &\ & 0&\  , \\
  &D_t\,\mathbbm{v}_z\ &\ + &\ c\partial_z\mathbbm{v}_0\ &\ &\  &\ =\ &\ -&2mc^2 & \mathbbm{t}_{1z}&\  , \\
  &D_t\,\mathbbm{t}_{1z}\ &\ &\ & + &\ 2cp_z\,\mathbbm{s}&\ =\ &\ &2mc^2 & \mathbbm{v}_z&\ ,
  \label{eqn:DHW12}
\end{alignat}
where
\begin{equation}
  D_t=\partial_t + q \int_{-1/2}^{1/2}{d\xi\,E_z(z+i\xi\partial_{p_z},t)\, \partial_{p_z}}
\end{equation}
is the reduced nonlocal operator and the vacuum initial conditions are
\begin{equation}\label{eqn:InitialConditions}
  \mathbbm{s}^\mathrm{vac}(p_z)=-\frac{2mc^2}{\omega(p_z)}\ , \quad
  \mathbbm{v}_z^\mathrm{vac}(p_z)=-\frac{2cp_z}{\omega(p_z)} \ ,
\end{equation}
with $\omega(p_z)=\sqrt{c^2p_z^2+m^2c^4}$. Furthermore, the $4\times4$ Wigner function in QED$_{1+1}$ can be further reduced to a $2\times2$ one, and the only changes are Eqs. (\ref{eqn:Expand1}) and (\ref{eqn:InitialConditions}). They become
\begin{equation}\label{eqn:Expand2}
  \mathcal{W}(z,p_z,t)=\frac{1}{2}(\mathbbm{1}\mathbbm{s}
+\sigma_z\mathbbm{v}_0+i\sigma_y\mathbbm{v}_z+i\sigma_x\mathbbm{t}_{1z}),
\end{equation}
and
\begin{equation}\label{eqn:InitialConditions1}
  \mathbbm{s}^\mathrm{vac}(p_z)=-\frac{mc^2}{\omega(p_z)}\ , \quad
  \mathbbm{v}_z^\mathrm{vac}(p_z)=-\frac{cp_z}{\omega(p_z)} \ .
\end{equation}

The particle number density in the coordinate space and the momentum space is defined as
\begin{equation}\label{eqn:NumDens}
\begin{split}
  f(z,t)&=\int \frac{dp_z}{2\pi}\,f(z,p_z,t)\, , \\
  f(p_z,t)&=\int dz\,f(z,p_z,t) \,
\end{split}
\end{equation}
with the phase space distribution function
\begin{equation}
\begin{split}
  f(z,p_z,t)\!=\!\tfrac{1}{\omega(p_z)}\{&mc^2[\mathbbm{s}(z,p_z,t)\!-\mathbbm{s}^\mathrm{vac}(p_z)]\\
  &+cp_z\,[\mathbbm{v}_z(z,p_z,t)\!-\mathbbm{v}_z^\mathrm{vac}(p_z)]\}.
\end{split}
\end{equation}
Note that this definition is based on the $2\times2$ Wigner function, see Eqs. (\ref{eqn:Expand2}) and (\ref{eqn:InitialConditions1}). If one prefers to use the $4\times4$ Wigner function, then the phase space distribution function should be divided by two.

The total number of created particles is
\begin{equation}
  N(t)=\int\frac{dp_z}{2\pi}\,f(p_z,t)=\int dz\,f(z,t).
\end{equation}

Finally, the particle number density and pair yield can be obtained by numerically solving Eqs. (\ref{eqn:DHW11})-(\ref{eqn:DHW12}) with the vacuum initial conditions Eq. (\ref{eqn:InitialConditions1}) employing spectral methods \cite{Boyd2001,Hesthaven2007,Kohlfurst2015}.
In addition, it should be noted that in order to compare with the CQFT where only one spin direction is considered, we divide the particle number density in momentum space and the total number of created particles by two in our final results, because the particle number density defined in Eq. (\ref{eqn:NumDens}) includes electrons as well as positrons.

\appendix


\begin{thebibliography}{99}\suppressfloats

\bibitem{Piazza2012}
A. Di Piazza, C. M\"{u}ller, K. Z. Hatsagortsyan, and C. H. Keitel, Rev. Mod. Phys. {\bf 84}, 1177 (2012).

\bibitem{Xie2017}
B. S. Xie, Z. L. Li, and S. Tang, Matter and Radiation at Extremes {\bf 2}, 225 (2017).

\bibitem{Dirac}
P. A. M. Dirac, Proc. R. Soc. A {\bf 117}, 612 (1928); ibid. {\bf 126}, 360 (1930).

\bibitem{Sauter}
F. Sauter, Z. Phys. {\bf 69}, 742 (1931); ibid. {\bf 73}, 547 (1932).

\bibitem{Heisenberg}
W. Heisenberg, H. Euler, Z. Phys. {\bf 98}, 714 (1936).

\bibitem{Schwinger}
J. Schwinger, Phys. Rev. {\bf 82}, 664 (1951).
\bibitem{Brezin1970}
E. Brezin and C. Itzykson, Phys. Rev. D {\bf 2}, 1191 (1970).

\bibitem{Popov1972}
V. S. Popov, Sov. Phys. JETP {\bf 34}, 709 (1972); ibid. {\bf 35}, 659 (1972).

\bibitem{Kim2007}
S. P. Kim and D. N. Page, Phys. Rev. D {\bf 75}, 045013 (2007).

\bibitem{Dumlu1011}
C. K. Dumlu and G. V. Dunne, Phys. Rev. Lett. {\bf 104}, 250402 (2010); Phys. Rev. D {\bf 83}, 065028 (2011).

\bibitem{Strobel2015}
E. Strobel and S. S. Xue, Phys. Rev. D {\bf 91}, 045016 (2015).

\bibitem{Oertel2019}
J. Oertel and R. Sch\"{u}tzhold, Phys. Rev. D {\bf 99}, 125014 (2019).

%


\bibitem{Dunne2005}
G. V. Dunne and C. Schubert, Phys. Rev. D {\bf 72}, 105004 (2005).

\bibitem{Dunne2006}
G. V. Dunne, Q. H. Wang, H. Gies, and C. Schubert, Phys. Rev. D {\bf 73}, 065028 (2006).

\bibitem{Dumlu2011}
C. K. Dumlu and G. V. Dunne, Phys. Rev. D {\bf 84}, 125023 (2011).

\bibitem{Ilderton2015}
A. Ilderton, G. Torgrimsson, and J. W{\aa}rdh, Phys. Rev. D {\bf 92}, 025009 (2015).

\bibitem{Schneider2018}
C. Schneider, G. Torgrimsson, and R. Sch\"{u}tzhold, Phys. Rev. D {\bf 98}, 085009 (2018).
\bibitem{Kluger1998}
Y. Kluger, E. Mottola, and J. M. Eisenberg, Phys. Rev. D {\bf 58}, 125015 (1998).

\bibitem{Schmidt1998}
S. M. Schmidt, D. Blaschke, G. Ropke, S. A. Smolyansky, A. V. Prozorkevich, and V. D. Toneev, Int. J. Mod. Phys. E {\bf 7}, 709 (1998).

\bibitem{Bloch1999}
J. C. R. Bloch, V. A. Mizerny, A. V. Prozorkevich, C. D. Roberts, S. M. Schmidt, S. A. Smolyansky, and D. V. Vinnik, Phys. Rev. D {\bf 60}, 116011 (1999).

\bibitem{Kohlfurst2014}
C. Kohlf\"{u}rst, H. Gies, and R. Alkofer, Phys. Rev. Lett. {\bf 112}, 050402 (2014).

\bibitem{Gong2020}
C. Gong, Z. L. Li, B. S. Xie, and Y. J. Li, Phys. Rev. D {\bf 101}, 016008 (2020).

\bibitem{Bialynicki1991}
I. Bialynicki-Birula, P. G\'{o}rnicki, and J. Rafelski, Phys. Rev. D {\bf 44}, 1825 (1991).

\bibitem{Hebenstreit2010}
F. Hebenstreit, R. Alkofer, and H. Gies, Phys. Rev. D {\bf 82}, 105026 (2010).

\bibitem{Hebenstreit2011}
F. Hebenstreit, R. Alkofer, and H. Gies, Phys. Rev. Lett. {\bf 107}, 180403 (2011).

\bibitem{ZLLi1517}
Z. L. Li, D. Lu, and B. S. Xie, Phys. Rev. D {\bf 92}, 085001 (2015); Z. L. Li, Y. J. Li, and B. S. Xie, Phys. Rev. D {\bf 96}, 076010 (2017).


\bibitem{Kohlfurst1819}
C. Kohlf\"{u}rst and R. Alkofer, Phys. Rev. D {\bf 97}, 036026 (2018); C. Kohlf\"{u}rst, Phys. Rev. D {\bf 99}, 096017 (2019).

\bibitem{Zlli2019}
Z. L. Li, B. S. Xie, and Y. J. Li, Phys. Rev. D {\bf 100}, 076018 (2019).

\bibitem{Krekora2004}
P. Krekora, Q. Su, and R. Grobe, Phys. Rev. Lett. {\bf 92}, 040406 (2004).

\bibitem{Krekora2006}
P. Krekora, Q. Su, and R. Grobe, Phys. Rev. A {\bf 73}, 022114 (2006).

\bibitem{QSu2012}
Q. Su, W. Su, Z. Q. Lv, M. Jiang, X. Lu, Z. M. Sheng, and R.Grobe, Phys. Rev. Lett. {\bf 109}, 253202 (2012).

\bibitem{Gong2018}
C. Gong, Z. L. Li, and Y. J. Li, Phys. Rev. A {\bf 98}, 043424 (2018).

\bibitem{QSu2019}
Q. Su and R. Grobe, Phys. Rev. Lett. \textbf{122}, 023603 (2019).

\bibitem{Dumlu2009}
C. K. Dumlu, Phys. Rev. D {\bf 79}, 065027 (2009).

\bibitem{Blinne2016}
A. Blinne and E. Strobel, Phys. Rev. D {\bf 93}, 025014 (2016).

\bibitem{Strobel2014}
E. Strobel and S. S. Xue, Nucl. Phys. B {\bf 886}, 1153 (2014).

\bibitem{SPKim2019}
S. P. Kim and D. N. Page, arXiv:1904.09749.

\bibitem{Unger2019}
J. Unger, S. Dong, R. Flores, Q. Su, and R. Grobe, Laser Phys. {\bf 29}, 065302 (2019).

\bibitem{ELI}
Extreme Light Infrastructure (ELI), \\ http://www.eli-beams.eu/.

\bibitem{XCELS}
Exawatt Center for Extreme Light Stidies (XCELS), http://xcels.iapras.ru/.

\bibitem{Schutzhold2008}
R. Sch\"{u}tzhold, H. Gies, and G. V. Dunne, Phys. Rev. Lett. {\bf 101}, 130404 (2008).

\bibitem{Bell2008}
A. R. Bell and J. G. Kirk, Phys. Rev. Lett. {\bf 101}, 200403 (2008).

\bibitem{Dunne2009}
G. V. Dunne, H. Gies, and R. Sch\"{u}tzhold, Phys. Rev. D {\bf 80}, 111301(R) (2009).

\bibitem{Piazza2009}
A. Di Piazza, E. L\"{o}tstedt, A. I. Milstein, and C. H. Keitel, Phys. Rev. Lett. {\bf 103}, 170403 (2009).

\bibitem{Bulanov2010}
S. S. Bulanov, V. D. Mur, N. B. Narozhny, J. Nees, and V. S. Popov, Phys. Rev. Lett. {\bf 104}, 220404 (2010).

\bibitem{Titov2012}
A. I. Titov, H. Takabe, B. K\"{a}mpfer, and A. Hosaka, Phys. Rev. Lett. {\bf 108}, 240406 (2012).

\bibitem{ZLLi2014}
Z. L. Li, D. Lu, B. S. Xie, L. B. Fu, J. Liu, and B. F. Shen, Phys. Rev. D {\bf 89}, 093011 (2014).

\bibitem{Torgrimsson1619}
G. Torgrimsson, J. Oertel, and R. Sch\"{u}tzhold, Phys. Rev. D {\bf 94}, 065035 (2016); G. Torgrimsson, Phys. Rev. D {\bf 99}, 096007 (2019).

\bibitem{Olugh2019}
O. Olugh, Z. L. Li, B. S. Xie, and R. Alkofer, Phys. Rev. D {\bf 99}, 036003 (2019).

\bibitem{Orthaber2011}
M. Orthaber, F. Hebenstreit, and R. Alkofer, Phys. Lett. B {\bf 698}, 80 (2011).

\bibitem{Nuriman2012}
A. Nuriman, B. S. Xie, Z. L. Li, and D. Sayipjamal, Phys. Lett. B {\bf 717}, 465 (2012).

\bibitem{Kohlfurst2013}
C. Kohlf\"{u}rst, M. Mitter, G. von Winckel, F. Hebenstreit, and R. Alkofer, Phys. Rev. D {\bf 88}, 045028 (2013).

\bibitem{Otto2015}
A. Otto, D. Seipt, D. Blaschke, S. A. Smolyansky, and B. K\"{a}mpfer, Phys. Rev. D {\bf 91}, 105018 (2015).

\bibitem{Jiang2012}
M. Jiang, W. Su, Z. Q. Lv, X. Lu, Y. J. Li, R. Grobe, and Q. Su, Phys. Rev. A {\bf 85}, 033408 (2012).

\bibitem{Schneider2016}
C. Schneider and R. Sch\"{u}tzhold, J. High Energy Phys. {\bf 02}, 164 (2016).

\bibitem{Ababekri2019}
M. Ababekri, B. S. Xie, and J. Zhang, Phys. Rev. D {\bf 100}, 016003 (2019).


\bibitem{Dong2020}
S. Dong, J. Unger, J. Bryan, Q. Su, and R. Grobe, Phys. Rev. E {\bf 101}, 013310 (2020).

\bibitem{Braun1999}
J. W. Braun, Q. Su, and R. Grobe, Phys. Rev. A {\bf 59}, 604 (1999).

\bibitem{Mocken2008}
G. R. Mocken and C. H. Keitel, J. Comput. Phys. {\bf 199}, 558 (2004); Comput. Phys. Commun. {\bf 178}, 868 (2008).

\bibitem{Blinne2014}
A. Blinne and H. Gies, Phys. Rev. D {\bf 89}, 085001 (2014).

\bibitem{Su2012}
W. Su, M. Jiang, Z. Q. Lv, Y. J. Li, Z. M. Sheng, R. Grobe, and Q. Su, Phys. Rev. A {\bf 86}, 013422 (2012).

\bibitem{Dabrowski2014}
R. Dabrowski and G. V. Dunne, Phys. Rev. D {\bf 90}, 025021 (2014).

\bibitem{Keldysh1964}
L. V. Keldysh, Zh. Eksp. Teor. Fiz. {\bf 47}, 1945 (1964) [Sov. Phys. JETP {\bf 20}, 1307 (1965)].

\bibitem{Abdukerim2013}
N. Abdukerim, Z. L. Li, and B. S. Xie, Phys. Lett. B {\bf 726}, 820 (2013).

\bibitem{Kohlfurst2015}
C. Kohlf\"{u}rst, PhD Thesis, arXiv:1512.06082.

\bibitem{Boyd2001}
J. P. Boyd, \textit{Chebyshev and Fourier Spectral Methods} (Dover Publications, Inc., New York, 2000).

\bibitem{Hesthaven2007}
J. Hesthaven, S. Gottlieb, and D. Gottlieb, \textit{Spectral Methods for Time-Dependent Problems} (Cambridge University Press, Cambridge, England, 2007).




%


%
%
%
%
%
%



\end{thebibliography}
\end{document}